\documentclass[structabstract]{aa}

\usepackage{amsmath}   
\usepackage{graphicx}
\usepackage{txfonts}
\usepackage{natbib}

\bibpunct{(}{)}{;}{a}{}{,} 

\usepackage[pdftitle={Tidal obliquity evolution of potentially habitable planets}, colorlinks = true, breaklinks = true, citecolor = blue, linkcolor = blue, urlcolor = blue, pdfauthor = {Ren\'{e} Heller}]{hyperref}

\begin{document}

\title{Tidal obliquity evolution of potentially habitable planets}


\author{R. Heller  \inst{1,2} \and J. Leconte \inst{3} \and R. Barnes \inst{4,5}}

\institute{
Astrophysikalisches Institut Potsdam (AIP), An der Sternwarte 16, 14482 Potsdam, Germany\\ \email{rheller@aip.de}
\and
Hamburger Sternwarte, Graduiertenkolleg 1351 ``Extrasolar Planets and their Host Stars'' of the Deutsche Forschungsgesellschaft 
\and
\'Ecole Normale Sup\'erieure de Lyon, 46 all\'ee d'Italie, F-69364 Lyon cedex 07, France; \\
Universit\'e Lyon 1, Villeurbanne, F-69622, France; CNRS, UMR 5574, Centre de Recherche Astrophysique de Lyon\\ \email{jeremy.leconte@ens-lyon.fr}
\and
University of Washington, Dept. of Astronomy, Seattle, WA 98195\\
\email{rory@astro.washington.edu}
\and
Virtual Planetary Laboratory, USA
}

\titlerunning{Tidal obliquity evolution of potentially habitable planets}

\authorrunning{Heller et al.}

\date{Received date / Accepted 10 January 2011}

\abstract
{Stellar insolation has been used as the main constraint on a planet's potential habitability. However, as more Earth-like planets are discovered around low-mass stars (LMSs), a re-examination of the role of tides on the habitability of exoplanets has begun. Those studies have yet to consider the misalignment between a planet's rotational axis and the orbital plane normal, i.e. the planetary obliquity.
}
{This paper considers the constraints on habitability arising from tidal processes due to the planet's spin orientation and rate. Since tidal processes are far from understood we seek to understand differences between commonly used tidal models.
}
{We apply two equilibrium tide theories -- a constant-phase-lag model and a constant-time-lag model -- to compute the obliquity evolution of terrestrial planets orbiting in the habitable zones around LMSs. The time for the obliquity to decrease from an Earth-like obliquity of $23.5^\circ$ to $5^\circ$, the `tilt erosion time', is compared to the traditional insolation habitable zone (IHZ) in the parameter space spanned by the semi-major axis $a$, the eccentricity $e$, and the stellar mass $M_\mathrm{s}$. We also compute tidal heating and equilibrium rotation caused by obliquity tides as further constraints on habitability. The Super-Earth Gl581\,d and the planet candidate Gl581\,g are studied as examples for these tidal processes.
}
{Earth-like obliquities of terrestrial planets in the IHZ around stars with masses $\lesssim~0.25\,M_\odot$ are eroded in less than 0.1\,Gyr. Only terrestrial planets orbiting stars with masses $\gtrsim~0.9\,M_\odot$ experience tilt erosion times larger than 1\,Gyr throughout the IHZ. Tilt erosion times for terrestrial planets in highly eccentric orbits inside the IHZ of solar-like stars can be $\lesssim~10$\,Gyr. Terrestrial planets in the IHZ of stars with masses $\lesssim~0.25\,M_\odot$ undergo significant tidal heating due to obliquity tides, whereas in the IHZ of stars with masses $\gtrsim~0.5\,M_\odot$ they require additional sources of heat to drive tectonic activity. The predictions of the two tidal models diverge significantly for $e~\gtrsim~0.3$. In our two-body simulations, Gl581\,d's obliquity is eroded to $0^\circ$ and its rotation period reached its equilibrium state of half its orbital period in $<~0.1$\,Gyr. Tidal surface heating on the putative Gl581\,g is $\lesssim~150$\,mW/m$^2$ as long as its eccentricity is smaller than 0.3.
}
{Obliquity tides modify the concept of the habitable zone. Tilt erosion of terrestrial planets orbiting LMSs should be included by atmospheric modelers. Tidal heating needs to be considered by geologists.
}


\maketitle

\section{Introduction}
\label{sec:introduction}

\subsection{The role of obliquity for a planetary atmosphere}
\label{sub:role_of_obliquity}

The obliquity of a planet, i.e. the angle $\psi_\mathrm{p}$ between its spin axis and the orbital normal, is a crucial parameter for the possible habitability of a planet. On Earth, the orbital angular momentum of the Moon stabilizes $\psi_\mathrm{p}$ at roughly $23.5^\circ$ against chaotic perturbations from the other solar system planets \citep{1993Natur.361..615L}. This steady tilt causes seasons and, together with the rotation period of 1\,d, assures a smooth temperature distribution over the whole globe with maximum variations of 150\,K. Measurements of oxygen isotope ratios of benthic foraminifera ($\delta^{18}O_\mathrm{b}$) in deep-sea sediments suggest that changes in the Earth's obliquity have caused phases of glaciations on a global scale \citep{Drysdale09182009}. In general, the coupled evolution of eccentricity ($e$) and obliquity has a fundamental impact on the global climate of terrestrial planets \citep{2010ApJ...721.1295D}, with higher obliquities rendering planets habitable at larger semi-major axes. \citet{1997Icar..129..254W} investigated varying obliquities for Earth and concluded that a substantial part of the Earth would not be tolerable for life if its obliquity were as high as 90$^\circ$. \citet{Hunt} studied the impact of zero obliquity on the temperature distribution on Earth's and finds a global contraction of the inhabitable area on Earth for such a case. Thus, both extremely high and very low obliquities pose a threat for a planet's habitability. A discussion of the prospects of eukaryotic life to survive a snowball Earth is given by \citet{Hoffmann}.

For $\psi_\mathrm{p}\lesssim~5^\circ$ the habitability of a terrestrial planet might crucially be hindered. Decreasing obliquities induce less seasonal variation of solar insolation between higher and lower latitudes. Thus, winters get milder and summers become cooler. Given that cool summer temperatures turn out to be more important than cold winters for the emergence of continental ice sheets, smaller tilt angles lead to more glaciation. As a consequence, the temperature contrast between polar and equatorial regions gets very strong \citep{2009ApJ...691..596S}, possibly leading to a collapse of the potential atmosphere (priv. comm. with Frank Selsis), which freezes out at the poles or evaporates at the equator.

\subsection{Variations and lockings of the obliquity}
\label{sub:variations}

It is generally assumed that planets form in a disk around the host star. It was thought that this disk is coplanar with the equatorial plane of the star and the planet's spin axis is aligned with the orbital normal. However, observational evidence points towards a more complex formation scenario, where the stellar obliquity, i.e. the misalignment between the \textit{stellar} rotation axis and the orbital plane normal, depends on stellar and planetary mass \citep{2010ApJ...718L.145W} as well as on the influence of perturbing bodies \citep{2007ApJ...669.1298F}. Furthermore, the spin axes of planets can experience significant re-orientations by giant impacts during the planetary formation period. Thus, the spin axes of the planets themselves are not necessarily perpendicular to the orbital plane. The obliquity $\psi_\mathrm{p}$ can have any orientation with values $0~\leq~\psi_\mathrm{p}~\leq~180^\circ$ \citep{1999Icar..142..219A,2001Icar..152..205C,2007ApJ...671.2082K,2010MNRAS.tmp..800M}, where a rotation prograde with the orbital motion means $0~\leq~\psi_\mathrm{p}~\leq~90^\circ$ and $\psi_\mathrm{p}~>~90^\circ$ defines a retrograde planet rotation.

In the course of the planet's lifetime, this possible spin-orbit misalignment can be subject to lockings or severe perturbations, e.g. by a third or more bodies inducing chaotic interaction \citep{1993CeMDA..57..293L, 1993Natur.361..615L} or Milankovi\'c cycles \citep{Milankovic,2010ApJ...721.1308S}; the putative presence of a moon that stabilizes the planet's obliquity as on Earth \citep{1997A&A...318..975N}; a perturber that pumps the planet into Cassini states \citep{1996Icar..122..166G} or drives the Kozai mechanism by leverage effects \citep{1962AJ.....67..591K,2007ApJ...669.1298F,2011MNRAS.411..565M}. In the case of Gl581\,d, e.g., orbital oscillations due to gravitational interactions occur with timescales of $10^3$ years \citep{2008AsBio...8..557B}.

Non-spherical, inhomogeneous planets may have their spin state modified by gravitational interactions with other bodies in the planetary system, assuming the system is not coplanar. These mass asymmetries provide a leverage that torques the planet. The amplitudes and frequencies of the variations are complex functions of the orbital evolution of the system as a whole, as the distances between all the planets, and hence the resulting gravitational torques, are constantly changing. In many cases, such as on Mars, the obliquity evolution is chaotic.

Eccentricity oscillations in the Gl581 system occur on $10^3$\,yr timescales \citep{2008AsBio...8..557B}. Thus, bodily obliquities will be significant if tidal processes between the planet and its host star are too weak to align the spin with the orbital plane normal on shorter time scales \citep{2010LPICo1538.5595B}. So far, the relative inclinations between most of the members of the Gl581 system are unknown, and hence the magnitudes of the torques cannot be estimated at present. Only the mutual inclination of $\upsilon$\,And\,A\,c and d has just been measured to be nearly $30^\circ$ \citep{2010ApJ...715.1203M}, suggesting that large variations may be possible among exoplanets. Moreover, the high eccentricity of Gl581\,d \citep{2009A&A...507..487M}, likely induced by planet-planet scattering \citep{2002Icar..156..570M}, suggests a significant inclination between the orbits of planet e through f. Therefore, we should expect that the obliquities of the Gl581 planets are more likely to be controlled by interactions among themselves, rather than tidal interactions with the star.

\subsection{Tidal effects induced by planetary obliquity}

Tidal interaction between a planet and its host star may alter the orbital and structural characteristics of the bodies. The evolution of the planet's semi-major axis $a$, eccentricity $e$, its rotation period $P_\mathrm{rot.}$, and $\psi_\mathrm{p}$ can put constraints on its habitability. In the long term, as discussed below, tidal effects reduce a planet's initial obliquity. We call this effect `\textit{tilt erosion}'. Given all the dependencies of a planet's climate on the body's obliquity, as described above, an investigation of tilt erosion is required.

Tidal processes imply exchange of orbital and/or rotational momentum between the bodies. The total angular momentum is conserved but friction in the objects transforms orbital energy into heat, which is released in the two bodies. This effect, called `\textit{tidal heating}', can alter the planet's geology and thus also constrains habitability. On the rocky moon Io e.g., surface heating rates of 2\,W/m$^2$ \citep{2000Sci...288.1198S}, excited by tidal distortions from Jupiter, are related to global volcanism. It is currently not known if a planet's geophysical activity is controlled by the surface heat flux \citep{1988Icar...75..187S,1991Icar...90..222S,2009epsc.conf..372S,2009epsc.conf..366P} -- or vice versa. Instead, tectonic activity on a planet may be determined by the volumetric heating rate or the degree of lithospheric fracturing or as the case may be, but these properties are currently not measurable remotely. While surface tidal heating rates may set an upper limit for a planet's habitability, they may also provide a lower limit. Here, we think of such planets, where tidal heating is the major source of inner energy that could drive geologic activity. In the solar system we find this constellation for various moons orbiting the giant planets, such as on Io and Europa around Jupiter, Enceladus orbiting Saturn, and Triton escorting Neptune.

The role of plate tectonics for the emergence and survival of life has been discussed, e.g. in \citet{2003ARA&A..41..429K} and \citet{2005AsBio...5..100G}. In the terrestrial planets of the solar system, as well as in the Moon, the radiogenic decay of long-lived isotopes $^{40}$~K, $^{232}$~Th, $^{235}$~U, and $^{238}$~U provided an energy source that drove or drives structural convection \citep{1991Icar...90..222S,2005AsBio...5..100G}. The current mean surface output of radioactive decay on Earth is $\approx~0.04\,\mathrm{W/m}^2$ while the global mean heat flow is $\approx~0.086\,\mathrm{W/m}^2$ \citep[see Sect. 6 and Fig.~8 in][]{2007SSRv..129...35Z}. While an Earth-sized object obviously can maintain its radiogenic heat source for several Gyr, the heat flow in a Mars-sized planet decreases much more rapidly. Early Mars tectonically froze when its surface heating rates, driven by radiogenic processes and not by tides, dropped below roughly 0.04\,W/m$^2$ \citep{1982JGR....87.9881F,1997Natur.385..234W}. Currently its surface rates due to radiogenic processes are $\approx~0.03$\,W/m$^2$ \citep{1991Icar...90..222S}. Obliquity tides may deliver a source of energy sufficient to drive tectonic mechanisms \citep{1997Natur.385..234W} of exoplanets near or in the IHZ of LMSs.

Besides tilt erosion, another threat for a planet's habitability emerges from `\textit{tidal equilibrium rotation}'. If zero eccentricity coincides with zero obliquity, the tidal equilibrium rotation period of a planet will equal its orbital period. In the reference system of the synchronously rotating planet, the host star is then static. This configuration could destabilize the planet's atmosphere, where one side is permanently heated and the other one dark. As we will show, synchronous rotation is only possible if $\sin(\psi_\mathrm{p})~=~0$.

\subsection{Obliquities and the insolation habitable zone}

Previous studies \citep{2008AsBio...8..557B,2008MNRAS.391..237J,2009ApJ...700L..30B, 2010ASPC..430..133B,2009ApJ...707.1000H} investigated the impact of tidal heating and orbital evolution on the habitability of exoplanets but not the effect of tilt erosion. We investigate here the timescales for tilt erosion of terrestrial planets orbiting main-sequence stars with masses as high as $1\,M_\odot$. M dwarf stars are possible hosts for habitable planets \citep{2007AsBio...7...30T}, they contribute by far the majority of stars in the solar neighborhood, and transit signatures of extrasolar planets can be distinguished most easily in front of small -- thus low-mass -- stars. We calculate tilt erosion times in the parameter space of $e$, $a$, and stellar mass $M_\mathrm{s}$. These regions are then compared with the traditional insolation habitable zone (IHZ). This zone is commonly thought of as the orbital distance between the planet and its host star, where stellar insolation is just adequate to allow the existence of liquid water on the planetary surface. While the first attempts of this concept \citep{1993Icar..101..108K} were primarily based on experience from the solar system, recent approaches include further effects such as orbital eccentricity, tidal evolution, and the atmosphere of the planet \citep{2007A&A...476.1373S,2008AsBio...8..557B,2009ApJ...691..596S}. In addition to the tidal obliquity evolution of potentially habitable planets, we also examine the amount of tidal heating induced by obliquities.

Our models applied below describe the interaction of two bodies: a star and a planet. We neglect relativistic effects and those arising from the presence of more bodies and assume tidal interaction of the planet with its host star as the dominant gravitational process. Our approach does not invoke the geological or rheological response of the body to tidal effects.

This article is structured as follows: After this introduction, we devote Sect. \ref{sec:methods} to the two models of equilibrium tide and we motivate our choice of parameter values. In Sect. \ref{sec:constraints} we describe the constraints that may arise from tidal obliquity processes on the habitability of terrestrial planets, i.e. tilt erosion, tidal heating, and equilibrium rotation. Section \ref{sec:results} is dedicated to the results, in particular to a comparison with the traditional IHZ, while we discuss the results in Sect. \ref{sec:discussion}. In Sect. \ref{sec:conclusions} we conclude.

\section{Methods}
\label{sec:methods}

\subsection{Parametrizing tidal dissipation}
\label{sub:tidal_dissipation}

\subsubsection{Equilibrium tide theories}
\label{subsub:equilibrium_tides}

Most current theories for the tidal evolution of celestial bodies are derived from the assumption of equilibrium tides, originally developed by \citet{1879_Darwin}. This model assumes that the gravitational potential of the tide raiser can be expressed as the sum of Legendre polynomials $P_l$ and it is assumed that the elongated equilibrium shape of the perturbed body is slightly misaligned with respect to the line which connects the two centers of masses. This misalignment is due to dissipative processes within the deformed body and it leads to a secular evolution of the orbital elements of the system. The actual shape of a tidally deformed body is difficult to calculate directly, as it is a function of both the gradient of the gravitational force across the body's diameter, as well as the body's internal structure and composition. Initial attempts to model tidal evolution therefore made the assumption that the shape of a body could be well-represented by a superposition of surface waves with different frequencies and amplitudes. This approach does a reasonable job of reproducing the measurable tidal effects on the Earth-Moon and Jupiter-Io systems, but its applicability to large eccentricities ($\gtrsim~0.3$) is dubious \citep[see e.g.][]{2008CeMDA.101..171F,2009ApJ...698L..42G}. This `lag-and-add' method consists of a Fourier decomposition of the tidal forcing, where small tidal phase lags $\varepsilon_l(\chi_l)$, or geometrical lags $\delta_l(\chi_l)=\varepsilon_l(\chi_l)/2$ \citep{1964RvGSP...2..467M}, are added for each frequency $\chi_l$. The key flaw of this theory emerges from the fact that the `lag-and-add' procedure of polynomials $P_l$ is only reasonable for a tight range of tidal frequencies \citep{2009ApJ...698L..42G}.

As long as the amplitude of tidal deformation of the body is small, deviations from the equilibrium shape are assumed to be proportional to the distorting force \citep{Love_1927}. By analogy with the driven harmonic oscillator, the tidal dissipation function $Q$ has been introduced, where

\begin{equation}\label{specificQ}
Q^{-1}(\chi_l)=\frac{\tan{\Big(}\varepsilon(\chi_l){\Big )}}{1-{\Bigg(}\displaystyle{\frac{\pi}{2}}-\varepsilon(\chi_l){\Bigg)} \tan{\Big(}\varepsilon(\chi_l){\Big)}} \approx \tan{\Big (} \varepsilon(\chi_l) {\Big )}
\end{equation}

\noindent
\citep{2009CeMDA.104..257E}. This parameter is a measure of the tidal energy dissipated in one cycle (See Appendix of \citet[][]{2009CeMDA.104..257E} for a complete discussion).

To this point, no assumptions have been made about the frequency dependence of the lag. It is hard to constrain empirically and depends on the structure and composition of the deformed body. For the Earth, seismic data argue for $Q\propto\chi_l^\alpha$, with $0.2~\lesssim~\alpha~\lesssim~0.4$ and yr$^{-1}~\lesssim~\chi_l~\lesssim~10^7$\,Hz \citep{2007JGRE..11212003E,2007AIPC..886..131E}. For gaseous planets and for stars, numerical simulations and analytical studies find a complex response spectrum, which is extremely sensitive to the size and depth of the convective and radiative layers \citep{2004ApJ...610..477O,2009ApJ...696.2054G}. To derive simple estimates of the effects of tidal dissipation on the orbital evolution, two main prescriptions for the lag of the equilibrium have been established over the years.

One of these two approaches is called the `constant phase lag' (CPL) model. Motivated by the slow variation of the dissipation function for the Earth, $\varepsilon_l$ can be considered as constant for each tidal frequency, thus $\varepsilon_l\propto\chi_l^0$ \citep{1955ZA.....36..245G,1966Icar....5..375G,1964RvGeo...2..661K,1999ARA&A..37..533P,2008CeMDA.101..171F}. On the one hand, this procedure seems to fairly reproduce the rheology of the Earth. On the other hand, it results in discontinuities when the tidal frequency is close to zero, which is the case for nearly synchronized exoplanets.

The other approach, termed `constant time lag' (CTL) model, introduces the relation $\varepsilon_l\propto\chi_l^1$, which is equivalent to a fixed time lag $\tau$ between the tidal bulge and the line connecting the centers of the two bodies under consideration \citep{1879_Darwin,1972Moon....5..206S,1979M&P....20..301M,1980M&P....23..185M,1981A&A....99..126H,1998ApJ...499..853E}. This choice coincides with the visco-elastic limit \citep{2009ApJ...698L..42G}. It does not produce discontinuities for vanishing tidal frequencies, as they occur for the CPL model in the case of synchronous rotation, and it allows for a complete analytical calculation of the tidal effects without any assumption on the eccentricity \citep{2010A&A...516A..64L}. This is particularly relevant for the study of exoplanets in close orbits, which often turn out to be highly eccentric. However, the key flaw in this case is that coupling between the modes has been ignored.

We use two of the most recent studies on equilibrium tides, one representing the family of CPL models \citep[][FM08 in the following]{2008CeMDA.101..171F} and one representing the family of CTL models \citep[][Lec10 in the following]{2010A&A...516A..64L}, to simulate the tidal evolution of terrestrial planets around LMSs. When comparing the values of the key tidal parameters in these two theories, i.e. $Q$ for the CPL model and $\tau$ for the CTL model, we recall that both models yield similar results in the limit of $\psi_\mathrm{p}~\rightarrow~0$ and $e~\rightarrow~0$. For a non-synchronously rotating object, the semi-diurnal tides dominate and Eq.~\eqref{specificQ} yields

\begin{equation}\label{QforJ}
Q^{-1}\approx 2\,|\omega-n| \,\tau
\end{equation}

\noindent
for the leading frequency $\chi_l=2\,|\omega-n|$. In the case of a nearly synchronous planet, annual tides dominate, thus $\chi_l=n$ and $Q^{-1}\approx n \,\tau$ (see Lec10 for a detailed discussion).

\subsubsection{Response in celestial bodies}
\label{subsub:response}

In this paper, particular attention is drawn to the evolution of the spin-orbit orientation and the rotation period. As test objects we choose an Earth-mass planet and a terrestrial Super-Earth with $M_\mathrm{p} = 10\,M_\mathrm{Earth}$, where $M_\mathrm{p}$ is the mass of the planet and $M_\mathrm{Earth}$ is the mass of the Earth. Terrestrial bodies of such masses may have a variety of compositions \citep{2010ApJ...715.1050B}, thus their response to tidal processes can vary. Against this background, we choose a value of 0.5 for the radius of gyration $r_{\mathrm{g,p}}=I_\mathrm{p}^{1/2}M_\mathrm{p}^{-1/2}R_\mathrm{p}^{-1}$ of the terrestrial planet, where $I_\mathrm{p}$ is the moment of inertia around the rotational axis. The relationship between $M_\mathrm{p}$ and the planetary radius $R_\mathrm{p}$ is taken as $R_\mathrm{p}~=~(M_\mathrm{p}/M_\mathrm{Earth})^{0.27}~\times~R_\mathrm{Earth}$ \citep{2007Icar..191..337S}. For both the Earth analog and the Super-Earth we choose a tidal dissipation value of $Q~=~100$, owing to the values for Earth given by \citet{2001GeoJI.144..471R} and \citet{2009ApJ...707.1000H}. Measurements of the Martian dissipation function $Q_\mathrm{Mars}~=~79.91\pm0.69$ \citep[][]{2007A&A...465.1075L}, as well as estimates for Mercury, where $Q_\mathrm{Mercury}~<~190$, Venus, with $Q_\mathrm{Venus}~<~17$ \citep{1966Icar....5..375G}, Io, implying $Q_\mathrm{Io}~<~100$ \citep{1979Sci...203..892P,1980LPI....11..871P,1988Icar...75..187S}, and the Moon, with $Q_\mathrm{Moon}~=~26.5\pm1$ \citep{1994Sci...265..482D}, indicate a similar order of magnitude for the tidal dissipation function of all terrestrial bodies. For Super-Earths at distances $<~1$\,AU to their host star, the atmospheres are assumed to be no more massive than $10^{-1}\,M_\mathrm{p}$ \citep[Fig.~6 in][]{2006ApJ...648..666R}, typically much less massive. Although thermal tides can have an effect on a planet's tidal evolution \citep[as shown for Venus by][]{1978Natur.275...37I}, we neglect their contribution to the tidal evolution of our test bodies for simplicity. Over the course of the numerical integrations we use a fixed value for $Q$. In real bodies, $Q$ is a function of the bodies' rigidity $\mu$, viscosity $\eta$, and temperature $T$ \citep{1988Icar...75..187S,1990Icar...83...39F}. A comprehensive tidal model would couple the orbital and structural evolution of the bodies since small perturbations in $T$ can result in large variations in $Q$ \citep{1988Icar...75..187S,2002ApJ...573..829M,2007JGRE..11212003E}. To estimate the impact on our results arising from uncertainties in $Q$, we will show an example where the tidal dissipation function for an Earth-mass planet is varied by a factor of two, to values of 50 and 200, and the dissipation value for the Super-Earth is varied by factor of 5, to values of 20 and 500. \citet{2009ApJ...707.1000H} sum up estimates for $Q$ and the Love number of degree 2, $k_2$, for planets in the solar system. They find $k_2=0.3$ and $Q~=~50$ the most reasonable choice for an Earth-like planet \citep[for the relationship between $Q$ and $k_2$ see][]{2009ApJ...707.1000H}. We take $Q_\mathrm{p}~=~100$ and $k_2~=~0.3$ \citep{1995geph.conf....1Y} for the terrestrial planets, consistent with our previous studies. With $Q'~=~3/2~\times~Q/k_2$ we thus have $Q'_\mathrm{p}~=~500$. For the CPL model of FM08 we assume the dynamical Love number of the $i$th body, $k_{\mathrm{d},i}$, to be equal to $k_{2,i}$. For the star we take $Q_\mathrm{s}~=~10^5$ \citep{2010A&A...514A..22H}.

The tidal time lag $\tau$ of the Earth with respect to the Moon as a tide raiser has been estimated to 638\,s \citep{1997A&A...318..975N} and $\approx~600$\,s \citep[][p. 562 therein]{1977RSPTA.287..545L}. We apply the more recent value of 638\,s to model the tidal evolution of the terrestrial planets with the CTL model and apply a fixed $\tau_\mathrm{s}~=~1/(Q_\mathrm{s}n_\mathrm{ini})$, where $n_\mathrm{ini}$ is the orbital mean motion at the beginning of the integration. Over the course of the integration, $\tau_\mathrm{p}$ and $\tau_\mathrm{s}$ are fixed. For those examples, where we vary $Q_\mathrm{p}$ in the CPL model, the tidal time lag will be adapted to $\tau_\mathrm{p}~=~638\,\mathrm{s}~\times~100/Q_\mathrm{p}$ since roughly $\tau_\mathrm{p}~\propto~1/Q_\mathrm{p}$.

Empirical values for $Q_\mathrm{p}$ and $\tau_\mathrm{p}$ are mostly found by backwards integrations of the secular motions of the respective bodies, where initial conditions need to be assumed \citep[examples for the derivation of $Q$ values in the Solar System are given by][]{1966Icar....5..375G}, or by precise measurements of the evolution of the orbital elements. There is no general conversion between $Q_\mathrm{p}$ and $\tau_\mathrm{p}$. Only if $e~=~0$ and $\psi_\mathrm{p}~=~0$, when merely a single tidal lag angle $\varepsilon_\mathrm{p}$ exists, then the approximation $Q_\mathrm{p}~\approx~1/\varepsilon_\mathrm{p}~\approx~1/(2|n-\omega_\mathrm{p}|\tau_\mathrm{p})$ applies -- as long as $n-\omega_\mathrm{p}$ remains unchanged. Hence, a dissipation value for an Earth-like planet of $Q_\mathrm{p}~=~100$ is not equivalent to a tidal time lag of 638\,s, so the results for the tidal evolution will intrinsically differ among the CPL and the CTL model. However, both choices are common for the respective model.

\subsection{Equilibrium tide with constant phase lag (CPL)}
\label{sub:cpl}

The reconsideration of Darwin's theory \citep{1879_Darwin,1880RSPT..171..713D} by FM08 is restricted to low eccentricities and obliquities. To compute the orbital evolution of the star-planet system self-consistently, we numerically integrate a set of six coupled differential equations for the eccentricity $e$, the semi-major axis $a$, the two rotational frequencies $\omega_i$ ($i \in \{\mathrm{s},\mathrm{p}\}$, where the subscripts `s' and `p' refer to the star and the planet, respectively), and the two obliquities $\psi_i$ given by

\begin{align} \label{equ:e_FM}
  \frac{\mathrm{d}e}{\mathrm{d}t} \ =& \ - \frac{ae}{8 G M_1 M_2} \sum_{i \, \neq \, j}Z'_i \left( 2\varepsilon_{0,i} - \frac{49}{2}\varepsilon_{1,i} + \frac{1}{2}\varepsilon_{2,i} + 3\varepsilon_{5,i} \right) \\
  \frac{\mathrm{d}a}{\mathrm{d}t} \ =& \ \frac{a^2}{4 G M_1 M_2} \sum_{i \, \neq \, j} Z'_i  \ {\Bigg (} 4\varepsilon_{0,i} + e^2{\Big [} -20\varepsilon_{0,i} + \frac{147}{2}\varepsilon_{1,i} + \nonumber\\
  & \ \frac{1}{2}\varepsilon_{2,i} - 3\varepsilon_{5,i} {\Big ]} -4\sin^2(\psi_i){\Big [}\varepsilon_{0,i}-\varepsilon_{8,i}{\Big ]}{\Bigg )} \\
  \frac{\mathrm{d}\omega_i}{\mathrm{d}t} \ =& \ - \frac{Z'_i}{8 M_i r_{\mathrm{g},i}^2 R_i^2 n} {\Bigg (}4\varepsilon_{0,i} + e^2{\Big [} -20\varepsilon_{0,i} + 49\varepsilon_{1,i} + \varepsilon_{2,i} {\Big ]} + \nonumber\\
  & \ 2\sin^2(\psi_i) {\Big [} -2\varepsilon_{0,i} + \varepsilon_{8,i} + \varepsilon_{9,i} {\Big ]} {\Bigg )} \\  
  \frac{\mathrm{d}\psi_i}{\mathrm{d}t} \ =& \ \frac{Z'_i \sin(\psi_i)}{4 M_i r_{\mathrm{g},i}^2 R_i^2 n \omega_i} {\Bigg (} {\Big [} 1-\xi_i {\Big ]}\varepsilon_{0,i} + {\Big [} 1+\xi_i {\Big ]}{\Big \{}\varepsilon_{8,i}-\varepsilon_{9,i}{\Big \}} {\Bigg )} \label{equ:psi_FM} \ .
\end{align}

\noindent
In these equations, $Z'_i$ stands for

\begin{equation}\label{equ:Zp}
 Z'_i \coloneqq 3 G^2 k_{\mathrm{d},i} M_j^2 (M_i+M_j) \frac{R_i^5}{a^9} \ \frac{1}{n Q_i} \ ,
\end{equation}

\noindent
$\xi_i \coloneqq r_{\mathrm{g},i}^2 R_i^2 \omega_i a n / (G M_j)$, $G$ is Newton's gravitational constant, $n$ is the orbital mean motion or orbital frequency, $M_i$ is the mass of the $i$th body, and $R_i$ its mean radius. The algebraic signs of the tidal phase lags are given by

\begin{align}\label{equ:epsilon}
\nonumber
\varepsilon_{0,i} & = \Sigma(2 \omega_i - 2 n)\\
\nonumber
\varepsilon_{1,i} & = \Sigma(2 \omega_i - 3 n)\\
\nonumber
\varepsilon_{2,i} & = \Sigma(2 \omega_i - n)\\
\nonumber
\varepsilon_{5,i} & = \Sigma(n)\\
\nonumber
\varepsilon_{8,i} & = \Sigma(\omega_i - 2 n)\\
\varepsilon_{9,i} & = \Sigma(\omega_i) \ ,\\
\nonumber
\end{align}

\noindent
with $\Sigma(x)$ the sign of any physical quantity $x$, thus $\Sigma(x)~=~+~1~\vee~-~1~\vee~0$. As given by Eq. \eqref{equ:psi_FM}, the evolution of $\psi_\mathrm{i}$ is not an explicit function of $e$. However, it is implicitly coupled to $e$ since $\mathrm{d}\psi_\mathrm{i}~=~\mathrm{d}\psi_\mathrm{i}(a,\omega_\mathrm{i})$, and $\mathrm{d}a$ as well as $\mathrm{d}\omega_\mathrm{i}$ are functions of $e$. Thus, for our model of coupled differential equations $\mathrm{d}\psi_\mathrm{i}~=~\mathrm{d}\psi_\mathrm{i}(a(e,t),\omega_\mathrm{i}(e,t))$.

Though the total angular momentum of the binary is conserved, tidal friction induces a conversion from kinetic and potential energy into heat, which is dissipated in the two bodies. The total energy that is dissipated within the perturbed body, i.e. the tidal heating, can be determined by summing the work done by tidal torques (Eqs. (48) and (49) in FM08). The change in orbital energy of the $i$th body induced by  the $j$th body is given by

\begin{align}\label{equ:E_orb_mod1}
\nonumber
\dot{E}_{\mathrm{orb.},i} = \ \frac{Z'_i}{8} \ \times \ {\Big (} \ &4 \varepsilon_{0,i} + e^2 {\Big [}-20 \varepsilon_{0,i} + \frac{147}{2} \varepsilon_{1,i} + \frac{1}{2} \varepsilon_{2,i}\\
&- 3 \varepsilon_{5,i} {\Big ]} - 4 \sin^2(\psi_i) \ {\Big [}\varepsilon_{0,i} - \varepsilon_{8,i}{\Big ]} \ {\Big )} \ ,
\end{align}

\noindent
where we define the outer braces as $\Gamma_{\mathrm{orb.},i}$. The change in rotational energy is deduced to be

\begin{align}\label{equ:E_rot_mod1}
\nonumber
\dot{E}_{\mathrm{rot.},i} = \ - \frac{Z'_i}{8} \frac{\omega_i}{n} \ \times \ {\Big (}& \ 4 \varepsilon_{0,i} + e^2 {\Big [}-20 \varepsilon_{0,i} + 49 \varepsilon_{1,i} + \varepsilon_{2,i}{\Big ]}\\
& + 2 \sin^2(\psi_i) \ {\Big [}- 2 \varepsilon_{0,i} + \varepsilon_{8,i} + \varepsilon_{9,i}{\Big ]} \ {\Big )} \ ,
\end{align}

\noindent
where we define the outer braces as $\Gamma_{\mathrm{rot.},i}$. The total energy released inside the body then is

\begin{equation}\label{equ:E_tid_FM08}
\dot{E}_{\mathrm{tid.},i} = - \ (\dot{E}_{\mathrm{orb.},i} + \dot{E}_{\mathrm{rot.},i}) > 0 \ .
\end{equation}

\noindent
If the net torque on the spin over an orbit is zero, then the body is said to be `tidally locked' and the equilibrium rotation rate is

\begin{equation} \label{equ:omega_equ_FM}
  \omega_\mathrm{p}^\mathrm{equ.} = n (1+9.5e^2)
\end{equation}

\noindent
\citep{1966AJ.....71....1G,1999ssd..book.....M,2008CeMDA.101..171F}, and Eq. \eqref{equ:E_tid_FM08} can be written as

\begin{equation} \label{equ:E_tid_FM08_equ}
  \dot{E}_{\mathrm{tid.,p}}^\mathrm{equ.} = \frac{Z'_\mathrm{p}}{8} {\Big [} (1+9.5e^2) \times \Gamma_{\mathrm{rot.},i} - \Gamma_{\mathrm{orb.},i} {\Big ]}
\end{equation}

\noindent
for a planet in equilibrium rotation. Tidal heating rates in the planet are thus a function of $e$, $a$, and $\psi_\mathrm{p}$. Assuming a certain surface heating rate $h_\mathrm{p}^\mathrm{equ.}~=~\dot{E}_{\mathrm{tid.,p}}^\mathrm{equ.}/(4 \pi R_\mathrm{p}^2)$ is given for a planet in equilibrium rotation, then Eq. \eqref{equ:E_tid_FM08_equ} can be solved for the corresponding tilt and becomes

\begin{equation}\label{equ:tilt_FM}
|\psi_\mathrm{p}^{\mathrm{equ.}}| = \arcsin\left(\sqrt{ \frac{ \frac{\displaystyle h_\mathrm{p}^\mathrm{equ.} 32 \pi R_\mathrm{p}^2 }{\displaystyle Z'_\mathrm{p}} - (1+9.5e^2) \ \mathcal{A} + \mathcal{C} }{2 \ \mathcal{B} \ (1+9.5e^2)+4 \ \mathcal{D}} }\right) \ ,
\end{equation}

\noindent
where

\begin{align} \nonumber
\mathcal{A} & \coloneqq 4 \varepsilon_{0,i} + e^2 {\Big [}-20 \varepsilon_\mathrm{0,p} + 49 \varepsilon_\mathrm{1,p} + \varepsilon_\mathrm{2,p} {\Big ]} \\ \nonumber
\mathcal{B} & \coloneqq -2 \varepsilon_\mathrm{0,p} + \varepsilon_\mathrm{8,p} + \varepsilon_\mathrm{9,p}\\  \nonumber
\mathcal{C} & \coloneqq 4 \varepsilon_\mathrm{0,p} + e^2 {\Big [}-20 \varepsilon_\mathrm{0,p} + \frac{147}{2} \varepsilon_\mathrm{1,p} + \frac{1}{2} \varepsilon_\mathrm{2,p} - 3\varepsilon_\mathrm{5,p} {\Big ]}\\
\mathcal{D} & \coloneqq {\Big [} \varepsilon_\mathrm{0,p} - \varepsilon_\mathrm{8,p} {\Big ]} \ .
\end{align}

\noindent
For $e~\gtrsim~0.324$ Eq. \eqref{equ:omega_equ_FM} yields $\omega_\mathrm{p}^\mathrm{equ.}~>~2n$. Thus, the signs of the phase lags $\varepsilon_{0,\mathrm{p}}$, $\varepsilon_{1,\mathrm{p}}$, $\varepsilon_{2,\mathrm{p}}$, $\varepsilon_{8,\mathrm{p}}$, and $\varepsilon_{9,\mathrm{p}}$ make a discrete jump to $+~1$. Hence $\mathcal{B}~=~0~=~\mathcal{D}$ and Eq. \eqref{equ:tilt_FM} has no analytic solution. Moreover, in this case all the terms depending on $\psi_\mathrm{p}$ vanish in Eq. \eqref{equ:E_tid_FM08} because $\varepsilon_{0,\mathrm{p}}~=~\varepsilon_{8,\mathrm{p}}$ and $2\varepsilon_{0,\mathrm{p}}~=~\varepsilon_{8,\mathrm{p}}+\varepsilon_{9,\mathrm{p}}$. Therefore the concept of maximum or minimum obliquity consistent with a given heating rate cannot apply. This situation is a direct result of the exclusion of higher order terms in the second order CPL model and is an important limitation of the low$e$-low$\psi$ approximations. If more frequencies were added, this degeneracy would arise at higher eccentricities.

\subsection{Equilibrium tide with constant time lag (CTL)}
\label{sub:ctl}

Lec10 extended the model presented by \citet{1981A&A....99..126H} to arbitrary eccentricities and obliquities. For convenience and to ease comparison with the CPL model, we rewrite their equations for the evolution of the orbital parameters as

\begin{align} \label{equ:e_Lec}
  \frac{\mathrm{d}e}{\mathrm{d}t} \ =& \ \frac{11 ae}{2 G M_1 M_2} \sum_{i \, \neq \, j}Z_i \left( \cos(\psi_i) \frac{f_4(e)}{\beta^{10}(e)}  \frac{\omega_i}{n} -\frac{18}{11} \frac{f_3(e)}{\beta^{13}(e)} \right) \\
  \frac{\mathrm{d}a}{\mathrm{d}t} \ =& \  \frac{2 a^2}{G M_1 M_2} \sum_{i \, \neq \, j} Z_i \left( \cos(\psi_i) \frac{f_2(e)}{\beta^{12}(e)} \frac{\omega_i}{n} - \frac{f_1(e)}{\beta^{15}(e)} \right) \\
  \frac{\mathrm{d}\omega_i}{\mathrm{d}t} \ =& \ \frac{Z_i}{2 M_i r_{\mathrm{g},i}^2 R_i^2 n} \left( 2 \cos(\psi_i) \frac{f_2(e)}{\beta^{12}(e)} - \left[ 1+\cos^2(\psi) \right] \frac{f_5(e)}{\beta^9(e)} \frac{\omega_i}{n}  \right) \\  
  \frac{\mathrm{d}\psi_i}{\mathrm{d}t} \ =& \ \frac{Z_i \sin(\psi_i)}{2 M_i r_{\mathrm{g},i}^2 R_i^2 n \omega_i}\left( \left[ \cos(\psi_i) - \frac{\xi_i}{ \beta} \right] \frac{f_5(e)}{\beta^9(e)} \frac{\omega_i}{n} - 2 \frac{f_2(e)}{\beta^{12}(e)} \right) \label{equ:psi_LEC}
\end{align}

\noindent
where

\begin{equation}\label{equ:Z}
 Z_i \coloneqq 3 G^2 k_{2,i} M_j^2 (M_i+M_j) \frac{R_i^5}{a^9} \ \tau_i \ ,
\end{equation}

\noindent
$k_{2,i}$ is the potential Love number of degree 2 of the $i$th body, and the extension functions in $e$ are

\begin{align}\label{equ:f_e}
\nonumber
\beta(e) & = \sqrt{1-e^2} \ ,\\
\nonumber
f_1(e) & = 1 + \frac{31}{2} e^2 + \frac{255}{8} e^4 + \frac{185}{16} e^6 + \frac{25}{64} e^8 \ ,\\
\nonumber
f_2(e) & = 1 + \frac{15}{2} e^2 + \frac{45}{8} e^4 \ \ + \frac{5}{16} e^6 \ ,\\
\nonumber
f_3(e) & = 1 + \frac{15}{4} e^2 + \frac{15}{8} e^4 \ \ + \frac{5}{64} e^6 \ ,\\
\nonumber
f_4(e) & = 1 + \frac{3}{2} e^2 \ \ + \frac{1}{8} e^4 \ ,\\
f_5(e) & = 1 + 3 e^2 \ \ \ + \frac{3}{8} e^4 \ ,\\
\nonumber
\end{align}

\noindent
following the nomenclature of \citet{1981A&A....99..126H}. When $e~=0$ all these extensions converge to 1. Furthermore, for $\tau_i~=~1/(n Q_i)$ one finds $Z'_i~=~Z_i$.

The tidal heating rate in the $i$th body is given by

\begin{align}\label{equ:heat} \nonumber 
\dot{E}_{\mathrm{tid.},i} = & \ Z_i {\Bigg (} \frac{f_1(e)}{\beta^{15}(e)} - 2 \frac{f_2(e)}{\beta^{12}(e)} \cos(\psi_i) \frac{\omega_i}{n} + \\
& \ {\Big [} \frac{1+\cos^2(\psi_i)}{2} {\Big ]} \frac{f_5(e)}{\beta^9(e)} \left\{\frac{\omega_i}{n} \right\}^2 {\Bigg )} \ .
\end{align}

\noindent
The equilibrium rotation rate is

\begin{equation}\label{equ:equ}
\omega_\mathrm{p}^{\mathrm{equ.}} = n \frac{f_2(e)}{\beta^3(e) f_5(e)} \frac{2 \cos(\psi_\mathrm{p})}{1+\cos^2(\psi_\mathrm{p})}
\end{equation}

\noindent
\citep[see also][]{2007A&A...462L...5L,2008Icar..193..637W}, and therefore Eq. (\ref{equ:heat}) can be written as

\begin{equation}\label{equ:heat_equ}
\dot{E}_{\mathrm{tid.},p}^{\mathrm{equ.}} = \frac{Z_\mathrm{p}}{\beta^{15}(e)} \left[ f_1(e) - \frac{f_2^2(e)}{f_5(e)} \frac{2 \cos^2(\psi_\mathrm{p})}{1+\cos^2(\psi_\mathrm{p})} \right]
\end{equation}

\noindent
for a planet in equilibrium rotation. This function of $\psi_\mathrm{p}$ has minima at $\psi_\mathrm{p}~=0^\circ~\vee~180^\circ$ and maximum at $\psi_\mathrm{p}~=~90^\circ$. The symmetry of tidal heating around $90^\circ$ simply mirrors the fact that a translation between ($\psi_\mathrm{p}$,$\omega_\mathrm{p}^\mathrm{equ.}$) and ($\pi - \psi_\mathrm{p}$,$-\omega_\mathrm{p}^\mathrm{equ.}$) results in identical physical states.

For a given surface heating rate $h_\mathrm{p}^\mathrm{equ.}$ of a planet in equilibrium rotation, Eq. \eqref{equ:heat_equ} becomes

\begin{equation}\label{equ:tilt_LEC}
|\psi_\mathrm{p}^{\mathrm{equ.}}| = \arccos\left(\left\{ \left[\displaystyle f_1(e) - \frac{\displaystyle 4 \pi R_\mathrm{p}^2 h_\mathrm{p}^\mathrm{equ.} \beta^{15}(e)}{\displaystyle Z_\mathrm{p}} \right]^{-1}\frac{\displaystyle 2 f_2^2(e)}{\displaystyle f_5(e)}-1 \right\}^{-1/2} \right)
\end{equation}

\noindent
for $4\pi\,R_\mathrm{p}^2~h_\mathrm{p}^\mathrm{equ.}/Z_\mathrm{p}~\geq~(f_1-2f_2^2/f_5)/\beta^{15}$.

\subsection{Comparison of CPL and CTL models}

Both models converge for a limiting case, namely when both $e~\rightarrow~0$ and $\psi~\rightarrow~0$. If we assume that tides have circularized the orbit and the rotation rate of the planet has been driven close to its equilibrium rotation, $\omega_\mathrm{p}~=~n$ in the CPL model\footnote{Recall Eq. \eqref{equ:equ} and see that $\omega_\mathrm{p}^\mathrm{equ.}$ also depends on $\psi_\mathrm{p}$ in the CTL model.}, then the signs of the tidal phase lags become $\varepsilon_{0,i}~=~0$, $\varepsilon_{1,i}~=~-~1$, $\varepsilon_{2,i}~=~1$, $\varepsilon_{5,i}~=~1$, $\varepsilon_{8,i}~=~-~1$, and $\varepsilon_{9,i}~=~1$. As an example consider the differential equation for the evolution of the obliquity. The other equations behave similarly. For $e~\rightarrow~0$ and $\psi_i~\rightarrow~0$, Eq. (\ref{equ:psi_FM}) transforms into

\begin{equation} \label{equ:psi_converge_FM}
  \frac{\mathrm{d}\psi_i}{\mathrm{d}t} \ = \ - \frac{Z'_i \ \psi_i}{2 M_i r_{\mathrm{g},i}^2 R_i^2 n \omega_i}   {\Big [} 1+\xi_i {\Big ]} ,
\end{equation}

\noindent
while Eq. (\ref{equ:psi_LEC}) from the CTL model at first order in $\psi$ and for $e~=~0$ gives

\begin{equation} \label{equ:psi_converge_LEC}
  \frac{\mathrm{d}\psi_i}{\mathrm{d}t} \ = - \ \frac{Z_i \ \psi_i}{2 M_i r_{\mathrm{g},i}^2 R_i^2 n \omega_i} \ \left[ 1 + \xi_i \right] .
\end{equation}

\noindent
Equation (\ref{equ:psi_converge_FM}) and Eq. (\ref{equ:psi_converge_LEC}) coincide for $\tau_i~=~1/(n Q_i)$, which is the value of the specific dissipation function for a quasi-circular pseudo-synchronous planet (see Sect. 3 in Lec10).

\section{Constraints on habitability from obliquity tides}
\label{sec:constraints}

\subsection{Tilt erosion}

We define the `tilt erosion time' $t_\mathrm{ero.}$ as the time required to reduce an initial Earth-like obliquity of 23.5$^\circ$ to 5$^\circ$. We numerically integrate the two sets of coupled equations from Sects. \ref{sub:cpl} and \ref{sub:ctl}, respectively, to derive $t_\mathrm{ero.}$ over a wide range of $e$, $a$, and $M_\mathrm{s}$. We choose two different planetary masses: an Earth-twin of one Earth mass and a Super-Earth of of 10 Earth masses. Planets are released with initial rotation periods of 1\,d orbiting the IHZs around stars with masses up to $1\,M_\odot$.

\subsection{Tidal heating from obliquity tides}

To calculate the obliquities $|\psi_\mathrm{min}^\mathrm{equ.}|$ and $|\psi_\mathrm{max}^\mathrm{equ.}|$ of a planet we use the adopted minimum and maximum surface heating rates of 0.04\,W/m$^2$ and 2\,W/m$^2$, respectively. Close to the star tidal heating will be $\gg~2$\,W/m$^2$ even without heating from obliquity tides, whereas in the outer regions obliquity tides may push the rates above the 2\,W/m$^2$ threshold or not -- depending on the actual obliquity. Further outside, there will be a minimum obliquity necessary to yield $h_\mathrm{p}^\mathrm{equ.}~=~0.04$\,W/m$^2$. We should bear in mind that these key heating rates of  0.04\,W/m$^2$ and 2\,W/m$^2$ are empirical examples taken from two special cases the Solar System. Depending on a planet's structure, composition, and age these thresholds may vary significantly. In general, the connection between surface heating rates and geologic activity is still subject to fundamental debate.

\subsection{Tidal locking vs. equilibrium rotation}
\label{sub:locking}

A widely spread misapprehension is that a tidally locked body permanently turns one side to its host \citep[e.g. in][]{1997A&A...318..975N,1997Icar..129..450J,2004A&A...425..753G,2007AsBio...7..167K}. Various other studies only include the impact of eccentricity on tidal locking, neglecting the contribution from obliquity \citep{1966Icar....5..375G,1966AJ.....71....1G,1998ApJ...499..853E,2000ApJ...537L..61T,2002A&A...385..166S,2004ApJ...610..464D,2007A&A...476.1373S,2008AsBio...8..557B}. As long as `tidal locking' denotes only the state of $\mathrm{d}\omega_\mathrm{p}/\mathrm{d}t~=~0$, the actual equilibrium rotation period, as predicted by the CTL model of Lec10, may differ from the orbital period, namely when $e~\neq~0$ and/or $\psi_\mathrm{p}~\neq~0$. Only if `tidal locking' depicts the recession of tidal processes in general, when $e~=~0$ and $\psi_\mathrm{p}~=~0$ in Lec10's model, then $\omega_\mathrm{p}~=~n$. As given by Eq. (\ref{equ:equ}), one side of the planet is permanently orientated towards the star if both $e~=~0$ and $\psi~=~0$ [\footnote{Structural inhomogeneities can also prevent rotation, as in the case of the Moon.}]. In this case, habitability of a planet can potentially be ruled out when the planet's atmosphere freezes out on the dark side and/or evaporates on the bright side \citep{1997Icar..129..450J}. As long as $e$ and $\psi_\mathrm{p}$ are not eroded, however, the planet can be prevented from an $\omega_\mathrm{p}~=~n$ locking. In the CPL model, however, the equilibrium rotation state is not a function of $\psi_\mathrm{p}$, thus `tidal locking', denoting $\mathrm{d}\omega_\mathrm{p}/\mathrm{d}t~=~0$, indeed can occur for $\psi_\mathrm{p}~\neq~0$.

In addition to atmospheric instabilities arising from $\omega_\mathrm{p}~=~n$, slow rotation may result in small intrinsic magnetic moments of the planet. This may result in little or no magnetospheric protection of planetary atmospheres from dense flows of coronal mass ejection plasma of the host star \citep{2007AsBio...7..167K} \citep[but see also][who make counterpoints]{2010arXiv1011.5798G,2011arXiv1101.0800G}. Again, since obliquities can prevent a synchronous rotation of the planet with the orbit, its magnetic shield can be maintained. An example of equilibrium rotation rates is shown in \citet{2010ASPC..430..133B}. Below, we apply the CPL and the CTL models to the Super-Earth Gl581\,d and explore its tidal equilibrium period as a function of $e$ and $\psi_\mathrm{p}$.

\subsection{Compatibility with the insolation habitable zone}

Our constraints on habitability from obliquity tides are embedded in the IHZ as presented by \citet{2008AsBio...8..557B}, who enhanced the model of \citet{2007A&A...476.1373S} to arbitrary eccentricities \citep{2002IJAsB...1...61W}.

The first forms of life on Earth required between 0.3 and 1.8\,Gyr to emerge \citep{2005AsBio...5..100G}. During this period, before the so-called Great Oxidation Event $\approx~2.5$\,Gyr ago \citep{2007Sci...317.1903A}, microorganisms may have played a major role in the evolution of Earth's atmosphere \citep{2002Sci...296.1066K}, probably enriching it with CH$_4$ (methane) and CO$_2$ (carbon dioxide). On geological times it was only recently that life conquered the land, about 450\,Myr ago in the Ordovician \citep{1997Natur.389..33K,2003Natur.42..248K}. Therefore it is reasonable to assume that a planet needs to provide habitable conditions for at least 1\,Gyr for life to imprint its spectroscopic signatures in the planet's atmosphere \citep{2002SPIE.4835...79S,2010MNRAS.407.1259J} or to leave photometrically detectable traces on the planet's surface \citep{2010ApJ...715..866F}. This is relevant from an observational point of view since the spectra of Earth-like planets will be accessible with upcoming space-based missions such as Darwin \citep{2006ESASP1306..505O,2007arXiv0707.3385L} and the Terrestrial Planet Finder \citep{2010AsBio..10..103K}. For this reason, when our two test objects in the IHZ show either $t_\mathrm{ero.}~\ll~1$\,Gyr or $h_\mathrm{p}^\mathrm{equ.}~\gg~2$\,W/m$^2$, constraints on the habitability of the planet will be appreciable.

\section{Results}
\label{sec:results}

\subsection{Time scales for tilt erosion}

\begin{figure*}
  \centering
  \scalebox{0.48}{\includegraphics{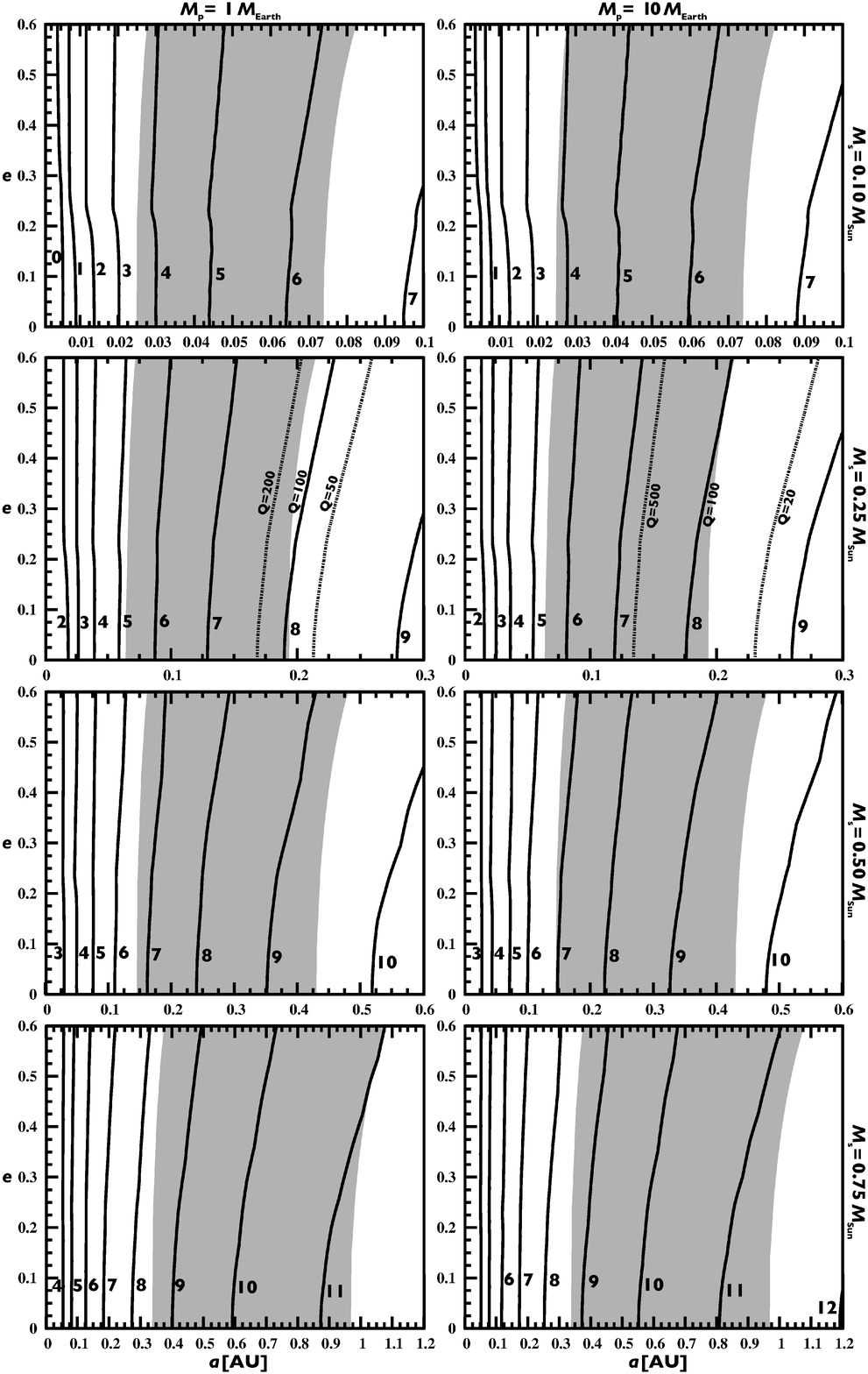}}
  \caption{Tilt erosion times for FM08. The IHZ is shaded in gray, contours of constant $t_\mathrm{ero.}$ are labeled in units of $\log(t_\mathrm{ero.}/\mathrm{yr})$. Error estimates for $Q_\mathrm{p}$ are shown for the 0.25\,$M_\odot$ star.}
  \label{fig:tilt_erosion_FM_ae}
\end{figure*}

\begin{figure*}
  \centering
  \scalebox{0.48}{\includegraphics{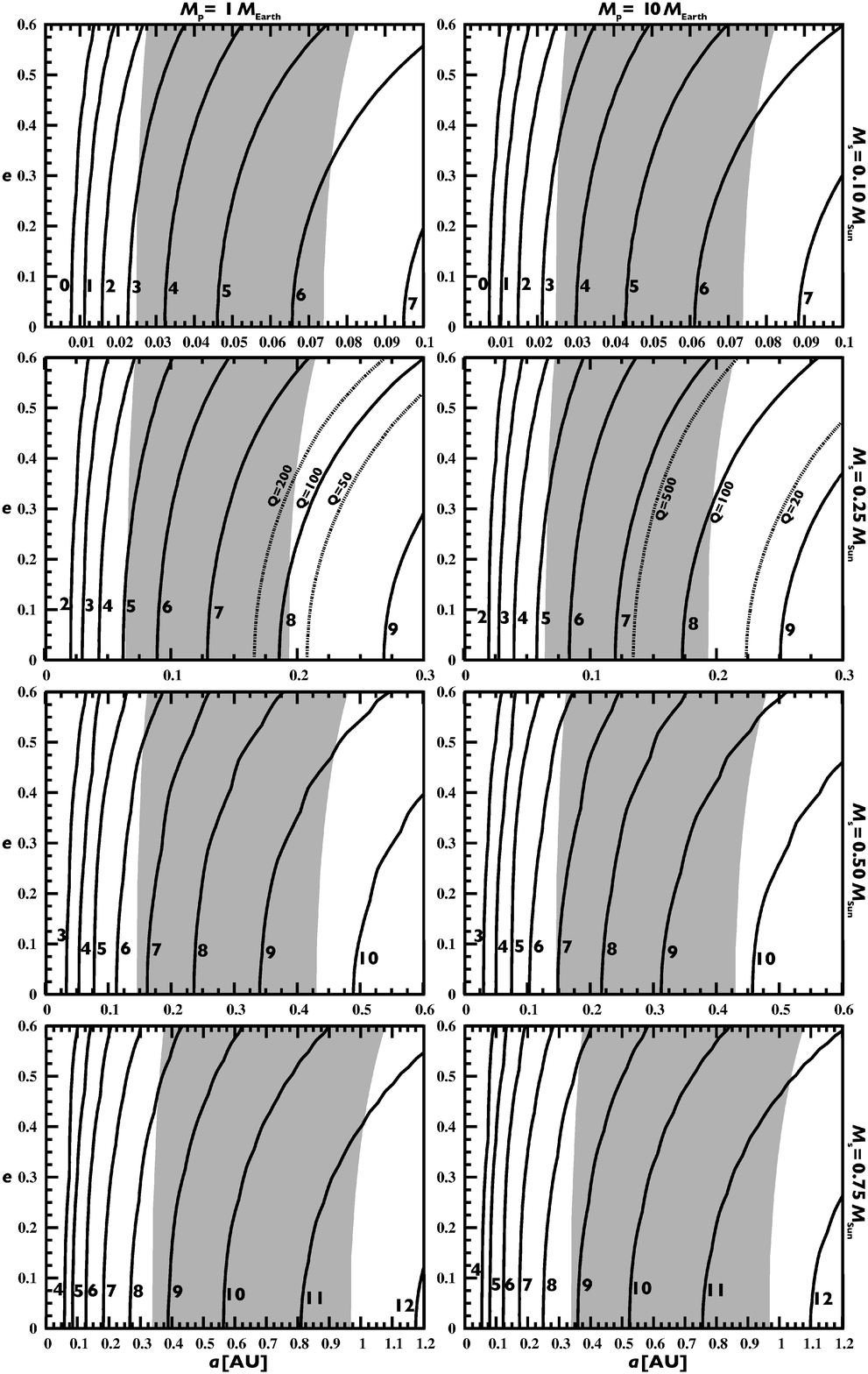}}
  \caption{Tilt erosion times for Lec10. The IHZ is shaded in gray, contours of constant $t_\mathrm{ero.}$ are labeled in units of $\log(t_\mathrm{ero.}/\mathrm{yr})$. Error estimates for $Q_\mathrm{p}$ are shown for the 0.25\,$M_\odot$ star.}  \label{fig:tilt_erosion_LEC_ae}
\end{figure*}

\begin{figure*}
  \centering
  \scalebox{0.42}{\includegraphics{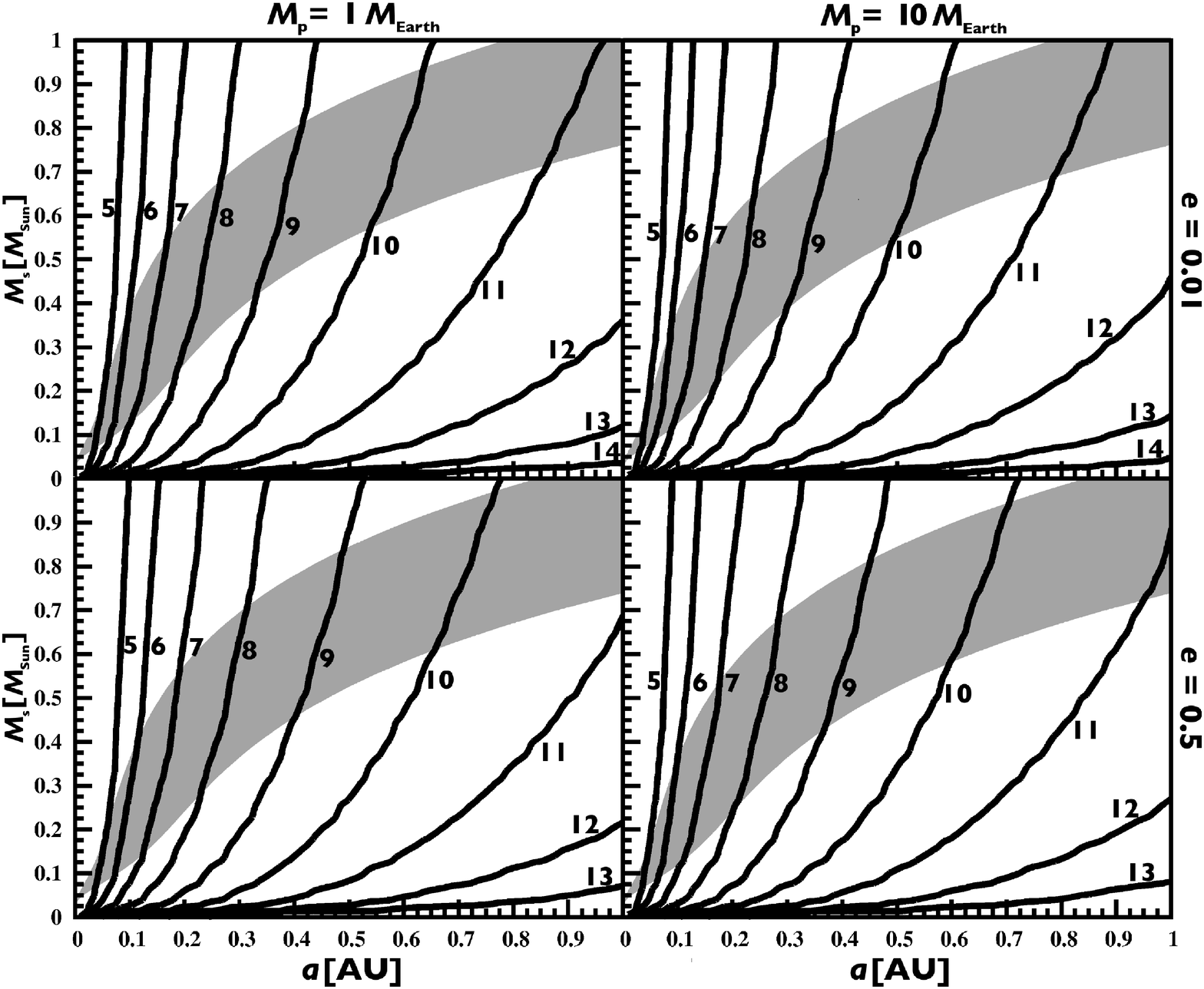}}
  \caption{Tilt erosion times for FM08. The IHZ is shaded in gray and the contours of constant tilt erosion times are labeled in units of $\log(t_\mathrm{ero.}/\mathrm{yr})$.}
  \label{fig:tilt_erosion_FM_Ma}
\end{figure*}

\begin{figure*}
  \centering
  \scalebox{0.42}{\includegraphics{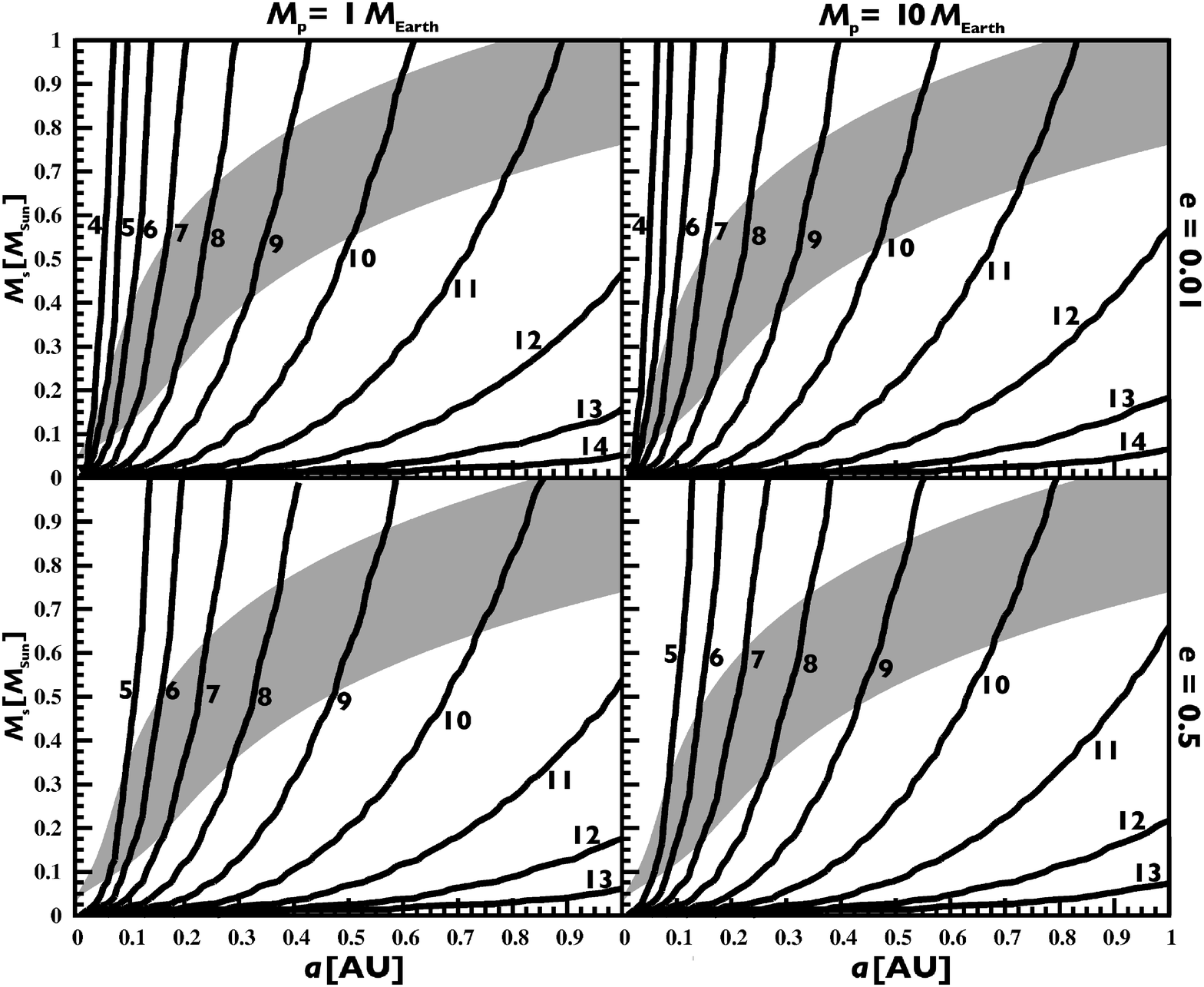}}
  \caption{Tilt erosion times for Lec10. The IHZ is shaded in gray and the contours of constant tilt erosion times are labeled in units of $\log(t_\mathrm{ero.}/\mathrm{yr})$.}  \label{fig:tilt_erosion_LEC_Ma}
\end{figure*}

We plot $t_\mathrm{ero.}$ for an Earth twin and a 10-Earth-mass planet orbiting a 0.1, 0.25, 0.5, and 0.75\,$M_\odot$ star in Fig.~\ref{fig:tilt_erosion_FM_ae} (CPL) and Fig.~\ref{fig:tilt_erosion_LEC_ae} (CTL). In Figs. \ref{fig:tilt_erosion_FM_Ma} and \ref{fig:tilt_erosion_LEC_Ma} we plot $t_\mathrm{ero.}$ for the two planets over a range of stellar masses and initial semi-major axes, for the two cases of initial eccentricities $e~=~0.01$ and $e~=~0.5$. We also include a few examples for different choices for $Q_\mathrm{p}$. The IHZs are shaded in grey. $t_\mathrm{ero.}$ increases with increasing initial distance from the star as well as with decreasing initial eccentricity. In eccentric orbits at short orbital distances the obliquity is eroded most rapidly.

For the FM08 model in Fig.~\ref{fig:tilt_erosion_FM_ae}, we see that $t_\mathrm{ero.}~\lesssim~0.1$\,Gyr for $M_\mathrm{s}~\lesssim~0.25\,M_\odot$ for both planets in the IHZ. For $M_\mathrm{s}~=~0.75\,M_\odot$ obliquities resist erosion for $\gtrsim~0.1$\,Gyr throughout the IHZ. Although we do not show the corresponding figures, we find that if the star has a mass $\gtrsim~1\,M_\odot$ the IHZ is completely covered with $t_\mathrm{ero.}~>~10$\,Gyr -- for all the three reasonable $Q$ values of both planets.

The calculations following Lec10 (Fig.~\ref{fig:tilt_erosion_LEC_ae}) yield qualitatively similar results for small eccentricities. However, for $e~\gtrsim~0.3$ their mismatch is of the same order of magnitude as the uncertainty in the tidal dissipation function $Q$ (see the $0.25\,M_\odot$ row). For these high eccentricities, the CTL model of Lec10 predicts significantly smaller tilt erosion times, where the discrepancy is also stronger for lower-mass stars.

A comparison of Figs.~\ref{fig:tilt_erosion_FM_ae} and \ref{fig:tilt_erosion_LEC_ae} at $e~=~0$ shows that, for a given combination of planetary and stellar mass, the contours for $t_\mathrm{ero.}~=~10^7$\,yr meet the abscissa at almost the same semi-major axis. For regions closer to the star, tilt erosion times provided by the CTL model are shorter. Farther outside, however, the CPL model yields smaller values for $t_\mathrm{ero.}$ at $e~=~0$. On the one hand, these discrepancies follow from the different forms of the differential equations -- remember that, although $e~=~0$, the initial obliquity $\psi_\mathrm{p}~=~23.5^\circ$ and thus the CPL and the CTL do \textit{not} converge as demonstrated in Eqs.~\eqref{equ:psi_converge_FM} and \eqref{equ:psi_converge_LEC} for $e~=~0~\wedge~\psi_\mathrm{p}=0$. On the other hand, the incompatible values for $Q_\mathrm{p}$ and $\tau_\mathrm{p}$ (see Sect. \ref{subsub:response}) lead to different evolutionary speeds in the two models.

The projection of $t_\mathrm{ero.}$ onto the $M_\mathrm{s}$-$a$ plane based on the FM08 model is shown in Fig.~\ref{fig:tilt_erosion_FM_Ma}. For the Earth twin in the $e~=~0.01$ case (upper left panel) we find that the complete IHZ is covered by $t_\mathrm{ero.}~>~1$\,Gyr as long as $M_s~\gtrsim~0.8\,M_\odot$, where the inner border of the IHZ is at roughly $0.4$\,AU. For an eccentricity of 0.5 (lower left panel) the Earth-sized planet requires an $M_s~\gtrsim~0.85\,M_\odot$ host star, thus a minimum semi-major axis of $0.5$\,AU, for tilt erosion times larger than 1\,Gyr throughout the IHZ. The larger moment of inertia of the $10\,M_\mathrm{Earth}$ Super-Earth (panels in the right column) assures that the initial obliquity of $23.5^\circ$ is washed out on timescales $>~1$\,Gyr even for slightly tighter orbits in the IHZ and somewhat lower-mass stars at both eccentricities. Comparing the upper and lower panel rows, corresponding to $e~=~0.01$ at the top and $e~=~0.5$ at the bottom, one sees that higher eccentricities tend to erode the spin faster by a factor of $~3$.

The higher order terms of $e$ in the equations of Lec10 produce a large discrepancy in $t_\mathrm{ero.}$ between $e~=~0.01$ (upper panel) and $e~=~0.5$ (lower panel) in Fig.~\ref{fig:tilt_erosion_LEC_Ma}. While the results for $e~=~0.01$ almost coincide with those of the FM08 model, Lec10 predicts much more rapid tilt erosion for $e~=~0.5$. The magnification of tidal effects at high eccentricities is due to the plummet of the mean orbital distance in eccentric orbits and is enhanced by the steep dependency of tidal effects on $a$ \citep{2008Icar..193..637W,2010A&A...516A..64L}. This behavior at high eccentricities is due to the slow convergence of elliptical expansion series and has been studied in detail by \citet{2010A&A...523A..87C}. For the lower left panel, where $e~=~0.5$ for the Earth-sized planet, an $M_\mathrm{s}\gtrsim~0.9\,M_\odot$ host is necessary for $t_\mathrm{ero.}~>~1$\,Gyr over the whole IHZ, with the inner border at about $0.55$\,AU. Again, the Super-Earth is slightly more resistive to tilt erosion. A compelling outcome of this visualization is that the Lec10 model predicts $t_\mathrm{ero.}~<~10$\,Gyr for an Earth-mass planet at the inner edge of the IHZ of a $1\,M_\odot$ star if the initial eccentricity is larger than $0.5$.

\subsection{Tidal heating from obliquity tides}
\label{sub:heating}

\begin{figure*}
\centering
  \scalebox{0.733}{\includegraphics{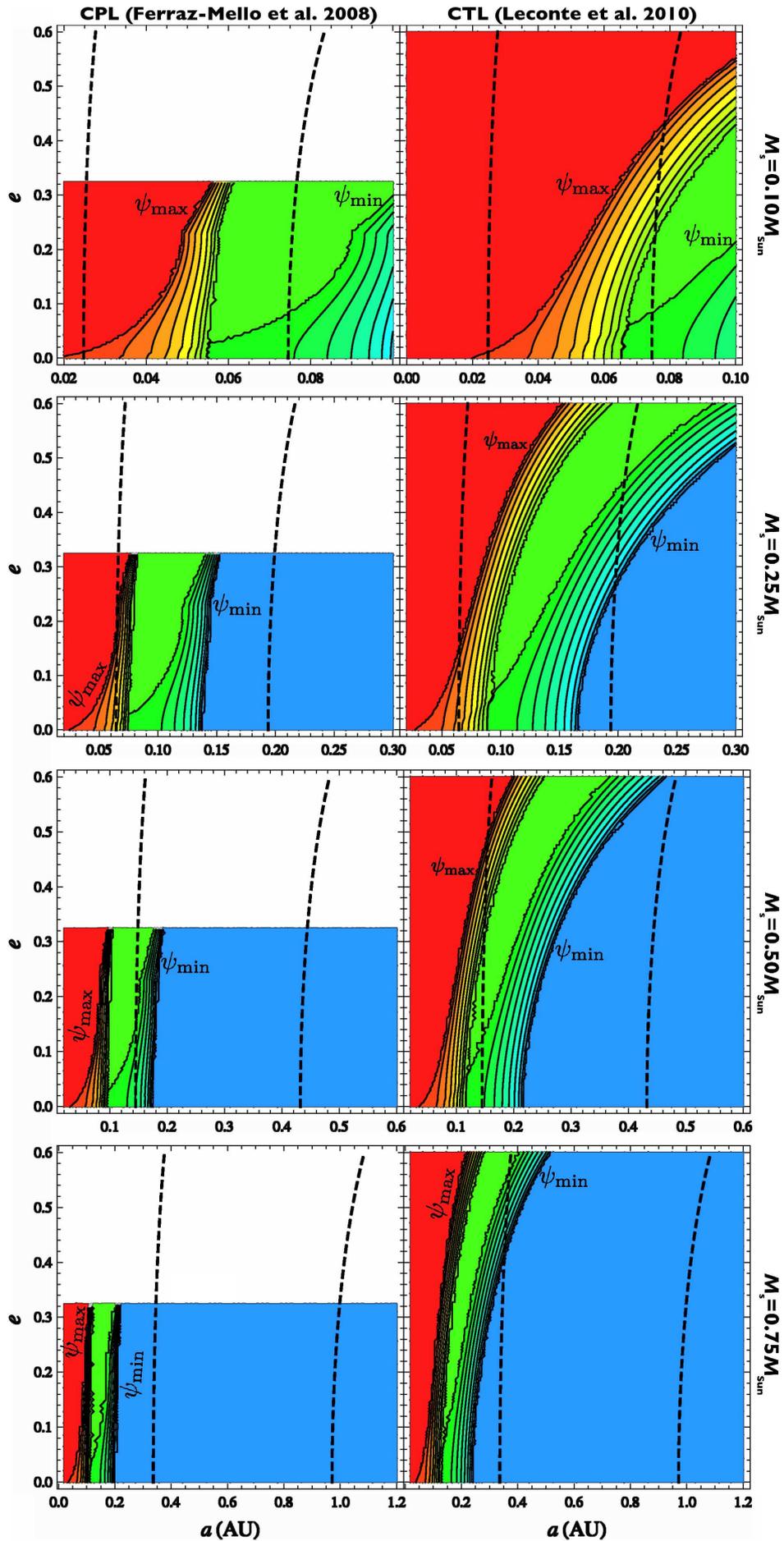}}
  \caption{Obliquity thresholds $\psi_\mathrm{max}$ and $\psi_\mathrm{min}$ as explained in the text. For each 9-tuple of contours the lines indicate $1^\circ, 20^\circ, 30^\circ, ..., 70^\circ, 80^\circ$, and $89^\circ$. The IHZ is indicated with dashed lines. In the red zone, $h_\mathrm{p}^\mathrm{equ.}~>~2\,\mathrm{W/m}^2$ for any obliquity. For $e~\gtrsim~0.327$ the CPL model breaks down, so this region is left blank.}
  \label{fig:psi}
\end{figure*}

In Fig.~\ref{fig:psi} we plot $\psi_\mathrm{max}^\mathrm{equ.}$ and $\psi_\mathrm{min}^\mathrm{equ.}$ as a function of $a$ and $e$ for an Earth-mass planet orbiting stars with four different masses, as above. Results for the $10\,M_\mathrm{Earth}$ terrestrial planet are not shown because they are very similar. The left column shows the results for the CPL model and the right column the CTL model. The white space in the left panels is due to the ambiguity of Eq. \eqref{equ:tilt_FM} as explained at the end of Sect. \ref{sub:cpl}. Red zones indicate that even for the case of $\psi_\mathrm{max}^\mathrm{equ.}~<~1^\circ$, tidal surface heating rates are $h_\mathrm{p}^\mathrm{equ.}>~2\,\mathrm{W/m}^2$. Starting from the left in a certain panel, the contours illustrate a maximum obliquity of $1^\circ, 20^\circ, 30^\circ, 40^\circ, 50^\circ, 60^\circ, 70^\circ, 80^\circ$, and $89^\circ$, all of which produce $h_\mathrm{p}^\mathrm{equ.}~=~2\,\mathrm{W/m}^2$ at their respective localization in the $a$-$e$ plane. Green in Fig.~\ref{fig:psi} depicts a region where neither any obliquity produces tidal surface heating rates $>~2\,\mathrm{W/m}^2$ nor any minimum obliquity is required to yield tidal surface heating rates $>~0.04\,\mathrm{W/m}^2$. The 9-tuple of contour lines for $\psi_\mathrm{min}^\mathrm{equ.}$ refers to obliquities of $1^\circ, 20^\circ, ..., 80^\circ$, and $89^\circ$ providing $h_\mathrm{p}^\mathrm{equ.}~=~0.04\,\mathrm{W/m}^2$ at their respective localization. Smaller obliquities lead to less tidal heating. In the blue zone finally, not even a planet with a rotational axis perpendicular to the orbital plane normal, e.g. $\psi_\mathrm{p}~=~90^\circ$, yields $h_\mathrm{p}^\mathrm{equ.}~\approx~0.04\,\mathrm{W/m}^2$.

Both models predict that Earth-like planets in the inner IHZ of stars with masses $\approx~0.1\,M_\odot$ are subject to intense tidal heating, while tidal heating in the outer zone of the IHZ is more moderate for the CPL model. With increasing stellar mass the IHZ is shifted to wider orbits for both models, whereas the green stripe of moderate tidal heating is located at roughly the same position in all the 8 panels of Fig.~\ref{fig:psi}. This is due to the weak sensitivity of $\psi_\mathrm{p}^\mathrm{equ.}$ on $M_\mathrm{s}$ compared to the sensitivity on $a$. The IHZ of a $0.25\,M_\odot$ star nicely covers the zone with moderate tidal heating. In this overlap, adequate stellar insolation meets tolerable tidal heating rates. Only highly eccentric orbits at the inner IHZ are subject to extreme tidal heating, as shown in the right column for the CTL model. Terrestrial planets in the IHZ of $0.5\,M_\odot$ stars do not undergo intense tidal heating. Only at the inner border of the IHZ $h_\mathrm{p}^\mathrm{equ.}~\approx~1\,\mathrm{W/m}^2$ for high obliquities, while at the outer regions tidal heating rates are of order $10\,\mathrm{mW/m}^2$ and smaller even for $\psi_\mathrm{p}~=~90^\circ$. For terrestrial planets in the IHZ of stars with masses $\gtrsim~0.75\,M_\odot$, both models indicate that tidal heating has negligible impact on the planet's energy budget and that it does not induce constraints on the planet's habitability.

\subsection{Gl581\,d and g as examples}

Gl581\,d is an $M_\mathrm{p}~\approx~5.6\,M_\mathrm{Earth}$ Super-Earth, grazing the outer rim of the IHZ of its $M_\mathrm{s}~\approx~0.29\,M_\odot$ host star \citep{2005A&A...443L..15B, 2007A&A...476.1373S, 2007A&A...476.1365V, 2008A&A...479..277B, 2009ApJ...700L..30B, 2009A&A...507..487M,2010A&A...522A..22W, 2010ApJ...723..954V}. The apoastron is situated outside the IHZ whereas the periastron is located inside. While the solar constant, i.e. the solar energy flux per time and area on the Earth's surface, is about 1000\,W/m$^2$ when the Sun is in the zenith\footnote{Outside of the Earth's atmosphere, i.e. at the distance of 1\,AU from the Sun, the solar constant is about 1400\,W/m$^2$.}, the stellar incident flux on Gl581\,d averaged over an orbit \citep{2002IJAsB...1...61W} is

\begin{equation}
f = \frac{L_\mathrm{Gl581}}{4 \pi a^2 \sqrt{1-e^2}} \approx 432\,\mathrm{W/m}^2 ,
\end{equation}

\noindent
with the luminosity of the host star $L_\mathrm{Gl581} = 0.013\,L_\odot$ \citep{2005A&A...442..635B,2009A&A...507..487M}. If the absorbed stellar flux per unit area is larger than $\approx~300$\,W/m$^2$, runaway greenhouse effects may turn a terrestrial planet uninhabitable \citep[][and references therein]{2007SSRv..129...35Z,2007A&A...476.1373S}. Thus, the maximum bond albedo for Gl581\,d compatible with this habitability criterion is $300/432~\approx~0.7$. A recent study on atmospheric constraints on the habitability of Gl581\,d is given by \citet{2010A&A...522A..23V}.

If confirmed, the recently detected planet candidate Gl581\,g \citep{2010ApJ...723..954V,2010arXiv1011.0186A} is located more suitably in the IHZ of Gl581, but it does not show a significant eccentricity. Since eccentricity has a crucial impact on a planet's orbital evolution, we choose Gl581\,d as an example to study the change of its spin orientation and period as well as its current tidal heating. An analysis of tidal heating in Gl581\,g is given at the end if this section. Since we take the parameters for Gl581\,d from \citet{2009A&A...507..487M} and the parameters for planet g from \citet{2010ApJ...723..954V}, our model for the two tidal evolutions need not necessarily be consistent.

\begin{figure*}
  \centering
  \scalebox{0.54}{\includegraphics{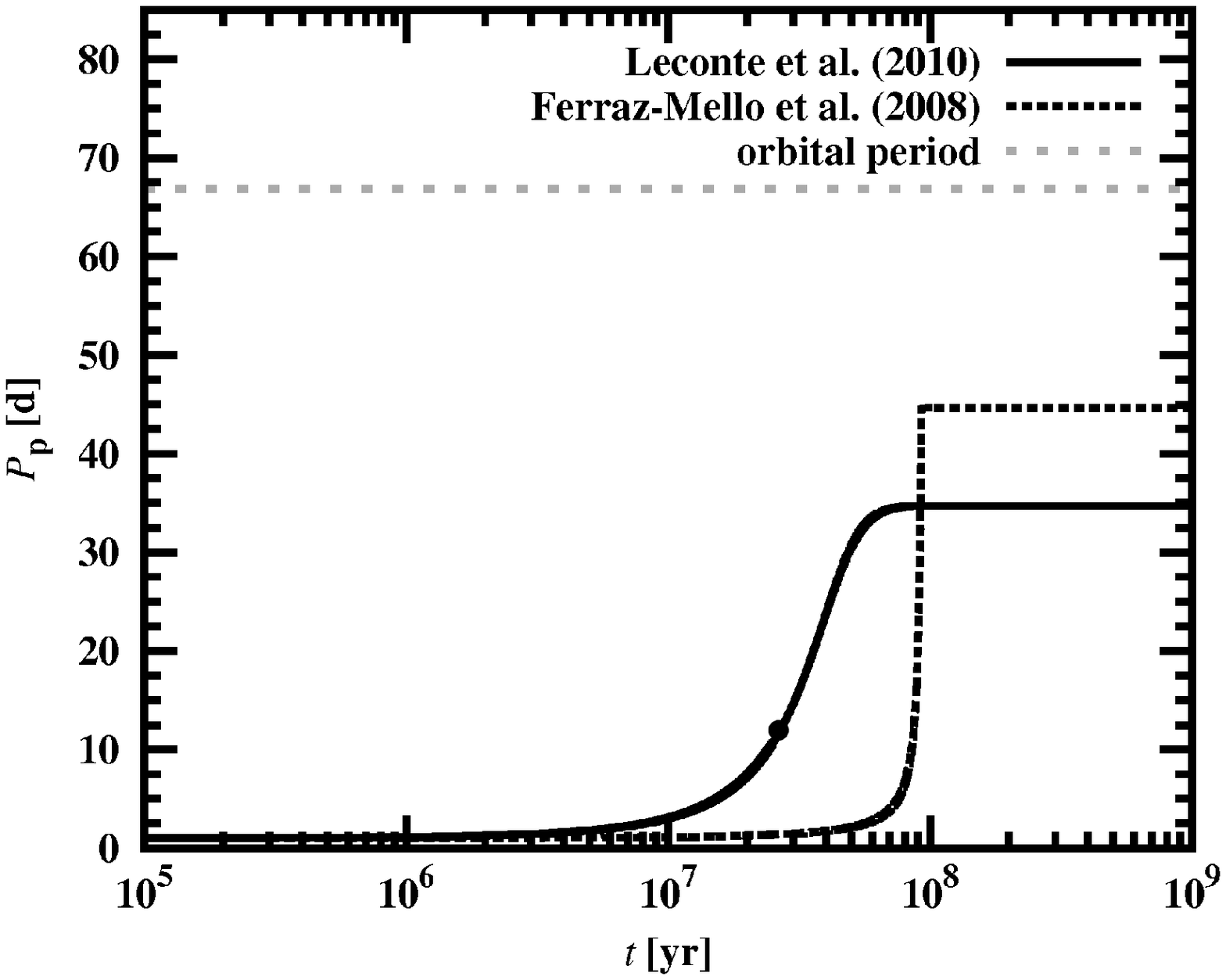}}
  \hspace{0cm}
  \scalebox{0.54}{\includegraphics{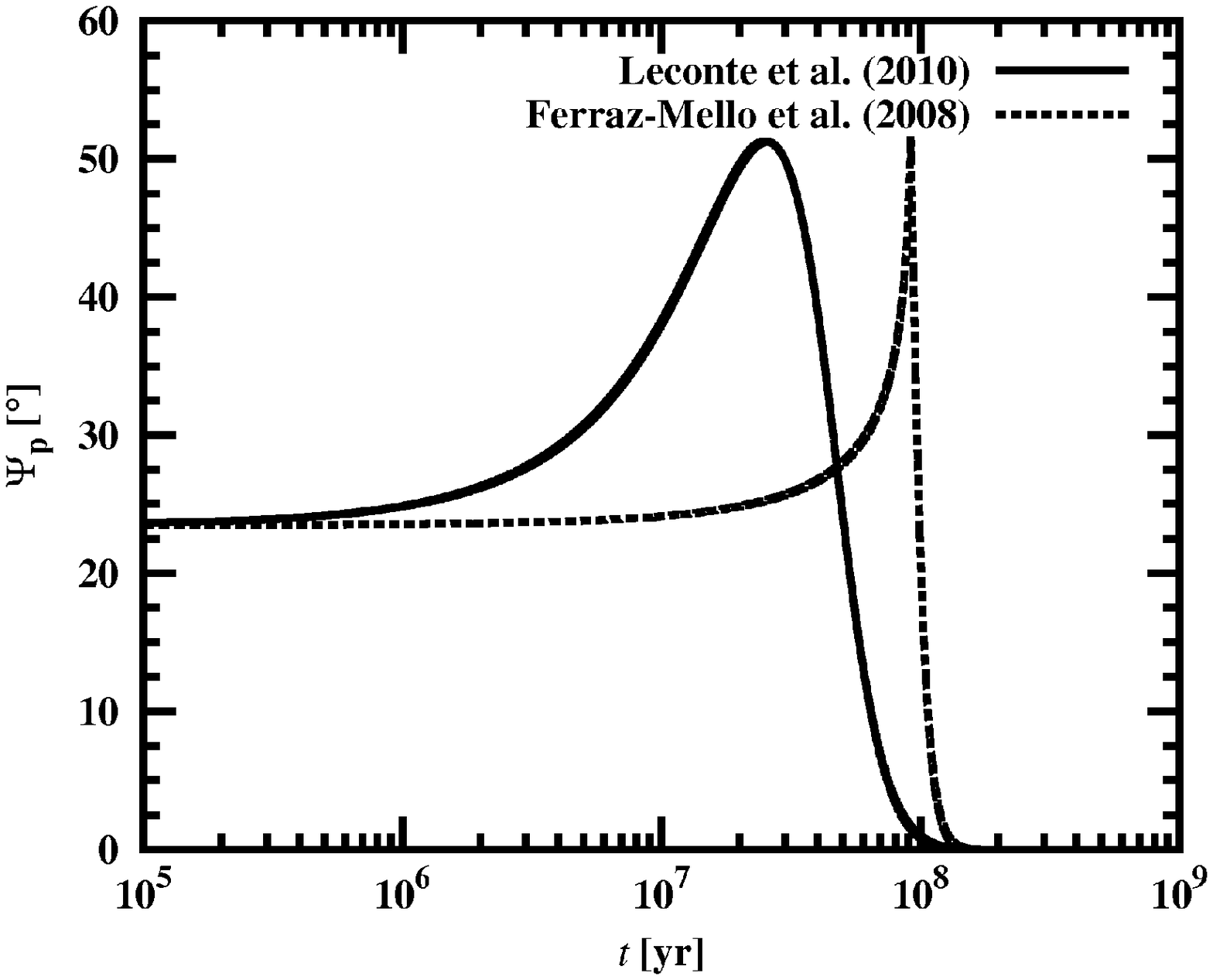}}
  \caption{\textit{Left:} Evolution of the putative rotation period for Gl581\,d for both models, FM08 and Lec10. The initial period was taken to be analog to the Earth's current day. The filled circle at $\approx~25.3$\,Myr on the Lec10 track indicates $\omega_\mathrm{p}/n~\lesssim~6.13$, thus from then on $\mathrm{d}\psi_\mathrm{p}/\mathrm{d}t~<~0$, as given by Eq. \eqref{equ:psi_sign}. \textit{Right:} Evolution of a putative initial Earth-analog obliquity for Gl581\,d for both models, FM08 and Lec10. }
  \label{fig:Gl581d_P_psi}
\end{figure*}

For the tidal evolution of Gl581\,d, a comparison between the FM08 and the Lec10 models is applied where it is reasonable. As initial obliquities we take $23.5^\circ$ for the planet and $0^\circ$ for the star, the initial rotation periods are $1$\,d for the planet and $94.2$\,d for the star \citep{2010ApJ...723..954V}. The remaining physical quantities of the host star and the two planets are assumed as explained in Sect. \ref{subsub:response}. For both models we find insignificant decreases in $e$ during the lifetime of the system, which has been estimated to $\gtrsim~2$\,Gyr \citep{2005A&A...443L..15B} and $\approx~4.3$\,Gyr \citep{2010ApJ...723..954V}. In the long term, FM08's model predicts a negligible increase of $a$, while Lec10's model yields an insignificant decrease. As shown in the left panel of Fig.~\ref{fig:Gl581d_P_psi}, both models predict a steady increase of the rotation period of Gl581\,d. For the model of Lec10, the numeric solution for the equilibrium rotation period of $P_\mathrm{equ.}~\approx~35$\,d can also be calculated analytically with Eq. (\ref{equ:equ}), using the observed eccentricity $e~=~0.38\,(\pm~0.09)$ and assuming that the spin axis is co-aligned with the orbit normal, $\psi_\mathrm{p}~=~0^\circ$. In the right panel of Fig.~\ref{fig:Gl581d_P_psi}, both models predict d's obliquity has been eroded over the course of the system's evolution, assuming no perturbations from other bodies. At an age of $\lesssim~0.1$\,Gyr, tidal processes would have pushed an Earth-like initial obliquity to zero and the rotation is most likely caught in its equilibrium state. In both models the obliquity of the planet first increases to $\approx~51^\circ$ before it is finally eroded. The reason for the change in the algebraic sign of $\mathrm{d}\psi_\mathrm{p}/\mathrm{d}t$ is found, when one sets $\mathrm{d}\psi_\mathrm{p}/\mathrm{d}t~>~0$. Considering that $\xi_\mathrm{p}~\ll1$ for Gl581\,d, $\mathrm{d}\psi_\mathrm{p}/\mathrm{d}t~>~0$ in Eq. \eqref{equ:psi_LEC} is equivalent to

\begin{equation} \label{equ:psi_sign}
\frac{\omega_\mathrm{p}}{n} \gtrsim \frac{2 f_2(e)}{f_5(e) \ \beta^3(e) \ \cos(\psi_\mathrm{p})} \ .
\end{equation}

\noindent
For $\psi_\mathrm{p}~=~51^\circ$ we find $\omega_\mathrm{p}/n~\gtrsim~6.13$. After about 25.3\,Myr in our simulations with the model of Lec10, the rotation has slowed down to $\omega_\mathrm{p}/n~\lesssim~6.13$ (see left panel in Fig.\ref{fig:Gl581d_P_psi}), thus $\mathrm{d}\psi_\mathrm{p}/\mathrm{d}t~<~0$. For the rest of the evolution $\mathrm{d}\psi_\mathrm{p}/\mathrm{d}t~<~0$ since the orbital period keeps increasing until it reaches the equilibrium state. Comparing both panels of Fig.~\ref{fig:Gl581d_P_psi}, we find that the time scale for the relaxation of the obliquity is longer than the time scale for the restoration of the equilibrium rotation period. This behavior has recently been explained by \citet{2009ApJ...704L...1C}.

\begin{figure*}
\centering
  \scalebox{0.58}{\includegraphics{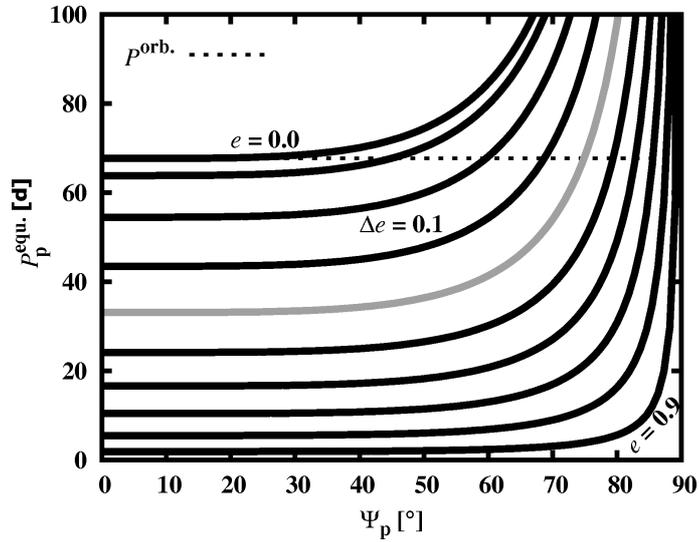}}
  \caption{Equilibrium rotation period of Gl581\,d as a function of obliquity for various eccentricities, following the model of Lec10. The gray line corresponds to $e~=~0.4$, close to the observed eccentricity of $e~=~0.38\,(\pm~0.09)$ \citep{2009A&A...507..487M}. The observed orbital period is indicated with a dotted line.}
  \label{fig:Gl581d_P_equ}
\end{figure*}

As given by Eq. (\ref{equ:equ}) from the CTL theory and as explained in Sect. \ref{sub:locking}, the equilibrium rotation period is also a function of $e$ and $\psi_\mathrm{p}$. We calculate the equilibrium rotation period $P_\mathrm{equ.}$ as a function of the obliquity $\psi_\mathrm{p}$ for a set of eccentricities (Fig.~\ref{fig:Gl581d_P_equ}). The most likely value of $P_\mathrm{equ.}~\approx~35$\,d, corresponding to $e~=~0.38$ and $\psi_\mathrm{p}~\lesssim~40^\circ$, can be inferred from the gray line in Fig.~\ref{fig:Gl581d_P_equ} for $e~=~0.4$. Since the observed orbital period of Gl581\,d is about $68$\,d, Fig.~\ref{fig:Gl581d_P_equ} shows that Gl581\,d does not permanently turn one hemisphere towards its host star, except for the cases of $\psi_\mathrm{p}~\approx~74^\circ~\vee~106^\circ$. {\b And even then the obliquity is so large that most of the surface cycles in and out of starlight over an orbit.}

We have done the same calculations of the tidal evolution for Gl581\,g, assuming a slight orbital eccentricity of 0.01 and a mass of $3.1\,M_\mathrm{Earth}$, corresponding to its projected minimum mass \citep{2010ApJ...723..954V}. In brief, we find an evolutionary scenario similar to that of planet d but on a considerably shorter timescale. After $\approx~20$\,Myr both tidal models predict a locking in equilibrium rotation at almost exactly the orbital period of $\approx~36.56$\,d. The evolution of $\psi_\mathrm{p}$ for planet g looks almost the same as for planet d, except that the obliquity is eroded about a factor of 5 faster in FM08's model and about a factor of 3 faster in the model of Lec10, due to the smaller planetary mass and semi-major axis of planet g.

\begin{figure*}
\centering
  \scalebox{0.58}{\includegraphics{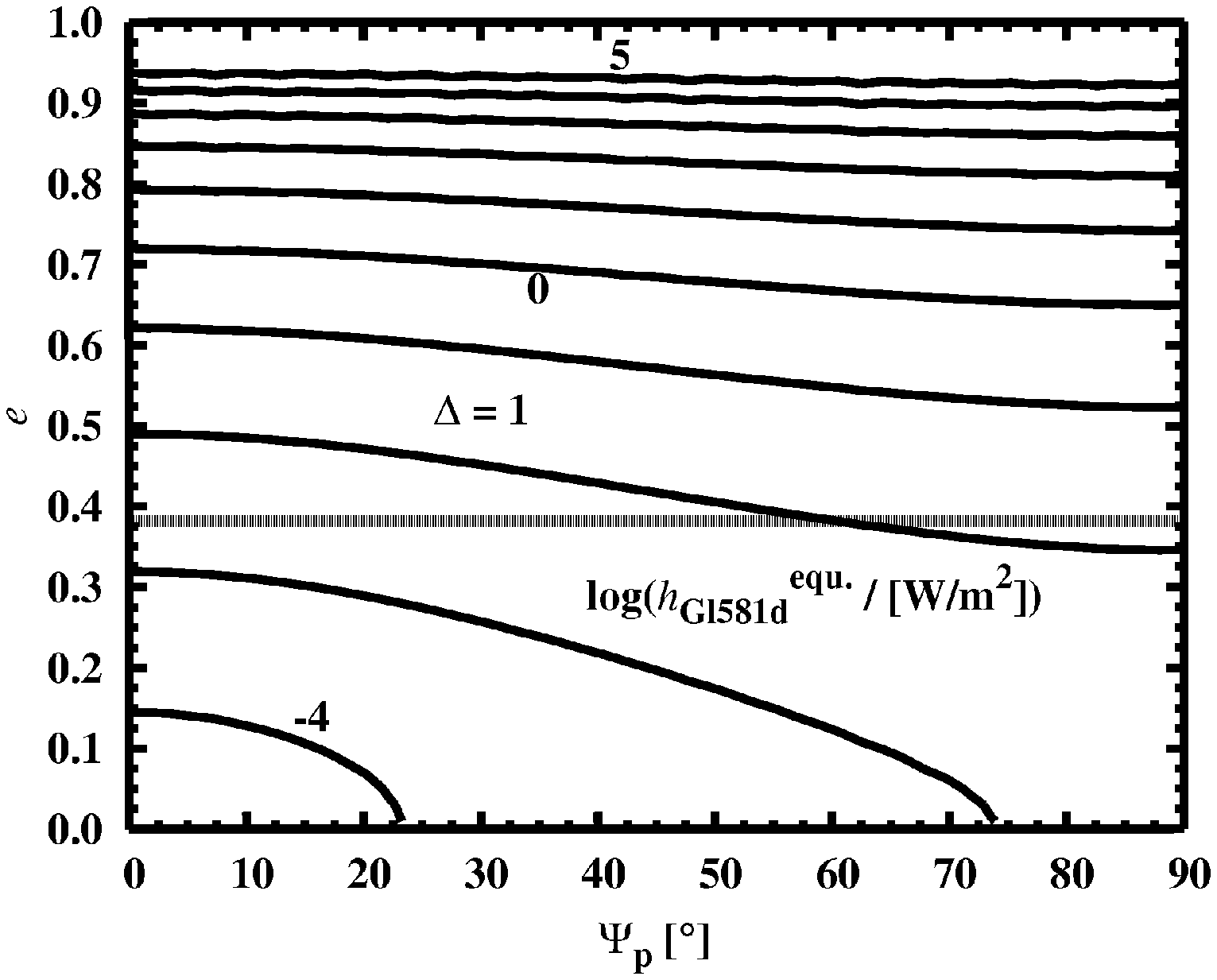}}
  \hspace{0cm}
  \scalebox{0.58}{\includegraphics{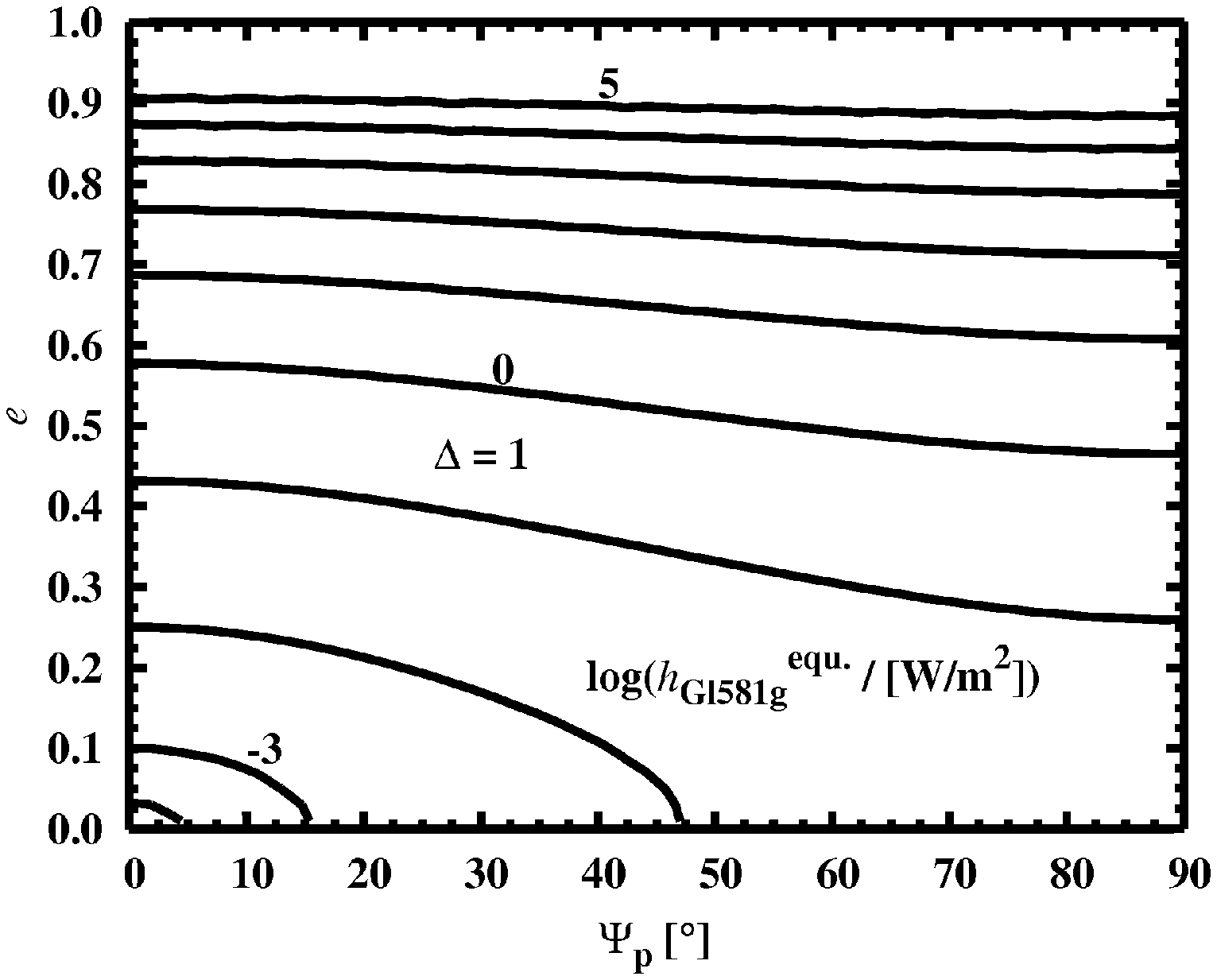}}
  \caption{Tidal surface heating rates, including eccentricity and obliquity heating, on Gl581\,d (left panel) and Gl581\,g (right panel) following the model of Lec10 as a projection on the parameter plane spanned by obliquity $\psi_\mathrm{p}$ and eccentricity $e$. Equilibrium rotation of the planets is assumed. The observed eccentricity of $0.38\,(\pm~0.09)$ for Gl581\,d \citep{2009A&A...507..487M} is indicated with a dashed line, while the eccentricity of Gl581\,g is compatible with 0. Depending on obliquity, tidal surface heating rates on Gl581\,d vary between roughly $2$ and $15$\,mW/m$^2$, while on planet candidate g they are $<~150$\,mW/m$^2$ as long as $e~\lesssim~0.3$.}
  \label{fig:Gl581dg_heat}
\end{figure*}

In Fig.~\ref{fig:Gl581dg_heat} we present the tidal surface heating rates, including obliquity heating, of Gl581\,d and g as a function of their putative obliquity. Since these planets encounter considerable interactions with the remaining four planets of the system, all orbits might be significantly tilted against each other (see Sect. \ref{sub:variations}). Thus, the obliquities $\psi_\mathrm{p}$ of the planets might be significant. Between spin-orbit alignment ($\psi_\mathrm{p}~=~0^\circ$) and a planetary spin axis perpendicular to the orbital plane ($\psi_\mathrm{p}~=~90^\circ$) tidal surface heating rates on Gl581\,d (left panel in Fig.~\ref{fig:Gl581dg_heat}) range from roughly $2$ to about $15$\,mW/m$^2$, given that $e~=~0.38\,(\pm~0.09)$. As a comparison, we show the tidal surface heating of the potentially habitable planet Gl581\,g in the right panel of Fig.~\ref{fig:Gl581dg_heat}. Measurements by \citet{2010ApJ...723..954V} indicate that the eccentricity of Gl581\,g is small if not zero. As long as its eccentricity is $<~0.3$ we find tidal heating on Gl581\,g to be $\lesssim~150\,\mathrm{mW/m}^2$. Only if the planet's eccentricity has been $\gtrsim~0.5$ in the history of the system, high obliquities could have produced tidal surface heating rates of $\approx~1$\,W/m$^2$ (upper half in the right panel of Fig.~\ref{fig:Gl581dg_heat}).

The eccentricities of the planets around Gl581 could be subject to fluctuations. In their history, their values might have been significantly larger than their current values. For reasonable eccentricities $<~0.7$ of Gl581\,d, we find that tidal surface heating rates on this planet have always been smaller than 1\,W/m$^2$ if the planet's semi-major axes has never been significantly smaller than its current value. On planet g tidal heating will be negligible if its eccentricity turns out to be $<~0.3$.

\section{Discussion}
\label{sec:discussion}

Our results for tilt erosion, tidal heating, and equilibrium rotation picture scenarios for atmospheric scientists and geophysicists. For low planetary obliquities, atmospheres of planets at the outer border of the IHZ can be subject to freeze out at the planet's poles (priv. comm. with Franck Selsis). Moreover, $\psi_\mathrm{p}~\rightarrow~0$ weakens the seasonal temperature contrast at a given location on the planetary surface but strengthens the temperature contrast as a function of latitude (see Sect.~\ref{sub:role_of_obliquity}). Given the mass and the age of a star, as well as the orbital parameters of the planet, our estimates of the tilt erosion time scales indicate whether a significant obliquity of the planet is likely. Our choice of $5^\circ$ as the final obliquity for the computation of $t_\mathrm{ero.}$ is somewhat arbitrary but the time for $\psi_\mathrm{p}$ to decline from $5^\circ$ to $0.5^\circ$ and even to $0.05^\circ$ is usually much smaller than $t_\mathrm{ero.}$ as we use it (see right panel in Fig.~\ref{fig:Gl581d_P_psi}). In any case, if low obliquities constrain the habitability of exoplanets, then the respective effects are likely to occur for $\psi_\mathrm{p}$ of the order of a few degrees and far before it has reached 0.

The spin angular momentum history of a planet, i.e. the evolution of obliquity and rotation rate, is a crucial feature of planetary climate. We encourage climate modelers to consider the diversity of atmospheres that are possible within the constraints outlined above.

Our exploration of tidal heating will be relevant for structural evolution models of terrestrial planets. Moreover, the tidal response is a function of the internal structure, hence a natural feedback mechanism exists between the two. The structure and cooling of Super-Earths have recently been explored in the literature \citep[e.g. by][]{2000JGR...10517563S,2007ApJ...670L..45V,2007GeoRL..3419204O,2009ApJ...700.1732K,2009epsc.conf..372S,2009epsc.conf..366P}, but a consensus has yet to emerge. On the one hand, the heat flow out of the interior should scale as the ratio of volume to the surface area, i.e. $\propto~R_i$. This relationship just assumes that larger planets have more radiogenic isotopes. Therefore large terrestrial planets seem more likely to have mobile lids than the Earth. On the other hand, larger planets have stronger gravitational fields and hence planets may fuse together due to the higher overburden pressure. Therefore large terrestrial planets seem more likely to have stagnant lids than the Earth. As if the story was not complicated enough already, none of those studies considered any potential tidal heating. Hence the issue for planets in the IHZ of LMSs is further complicated, but at least appears ripe for observational constraints: As recently shown by \citet{2010AJ....140.1370K}, acute volcanism on Earth-like planets will be detectable with the James Webb Space Telescope. Our calculations suggest that such worlds orbiting $\lesssim~0.25\,M_\odot$ stars orbiting closer than 0.1\,AU will present promising targets.

Tidal heating rates, as given by the models of Lec10 and FM08, are a function of $e$ and $\psi_\mathrm{p}$, amongst others. For Gl581\,d we find $2\,\mathrm{mW/m}^2~\lesssim~h_\mathrm{Gl581\,d}^\mathrm{equ.}(\psi_\mathrm{p})~\lesssim~15\,\mathrm{mW/m}^2$ for its current eccentricity. This is in line with the value of $0.01\,\mathrm{W/m}^2$ estimated by \citet{2009ApJ...700L..30B}, who neglected the planet's obliquity. We also explore $h_\mathrm{Gl581\,d}^\mathrm{equ.}$ as a function of the planet's eccentricity and find that tidal heating is unlikely to have constrained the habitability of the planet since $h_\mathrm{Gl581\,d}^\mathrm{equ.}~\lesssim~0.1\,\mathrm{W/m}^2$ as long as $e~<~0.7$. Tidal heating on the exoplanet candidate Gl581\,g, which would orbit its host star in the IHZ, is $\lesssim~150\,\mathrm{mW/m}^2$ as long as its eccentricity is smaller than 0.3.

\citet{2007A&A...476.1373S} approximated the rotational relaxation time of the planet to be $10$\,Myr, concluding that the planet has one permanent day side and night side. Indeed, with the revised value of $e~=~0.38\,(\pm~0.09)$ \citep{2009A&A...507..487M} we find that the tidal equilibrium rotation occurs after approximately 80\,Myr (Lec10) - 90\,Myr (FM). However, the equilibrium rotation period of the planet, as calculated with Lec10's ansatz, turns out to be $P_\mathrm{Gl581\,d}^\mathrm{equ.}~\approx~35\,\mathrm{d}~\approx~P_\mathrm{orb.}/2$. \citet{2008A&A...488L..63C} have approached the issue of equilibrium rotation of terrestrial planets for low eccentricities and obliquities. For Gl581\,d they found $\omega_\mathrm{Gl581\,d}^\mathrm{equ.}/n~\approx~1.25 \Leftrightarrow P_\mathrm{Gl581\,d}^\mathrm{equ.}~\approx~54.4\,\mathrm{d}$.

If the eccentricity of Gl581\,g turns out to be 0 and if this state was stable at least over the recent 1\,Myr, then tides have driven this planet into an equilibrium rotation state, where both $\psi_\mathrm{p}~=~0$ and $e~=~0$. Due to the smaller planetary mass and semi-major axis, as compared to planet d, Gl581\,g is driven into its equilibrium state faster by a factor between 3 (FM08) and 5 {Lec10} in $t_\mathrm{ero.}$.

The semantics of the terms `tidal locking' and `synchronous rotation' carry the risk to cause confusion. \citet{Wittgenstein_PU} has pointed out that the meaning of a word is given by its usage in the language (see \S 30, 43, and 432 therein). However, `tidal locking'  and `synchronous rotation' are used ambiguously in the literature (see Sect. \ref{sub:locking}), thus their meanings remain diffuse. While some authors simply refer to `tidal locking' as the state where a planet permanently turns one hemisphere towards its host star, others use it in a more general sense of tidal equilibrium rotation. Furthermore, the term `pseudo-synchronous rotation' appears, depicting $\mathrm{d}\omega_\mathrm{p}/\mathrm{d}t~=~0$ while $\omega_\mathrm{p}~\neq~n$. We caution that in the CTL model of Lec10 tidal processes drive the planetary rotation period towards the orbital period only if $e~=~0$ and $\psi_\mathrm{p}~=~0$. If a planet with $e~\neq~0$ or $\psi_\mathrm{p}~\neq~0$ is said to be tidally locked, then it is clear that tidal locking does not depict the state of the planet turning one hemisphere permanently to its star. However, for certain combinations of $e~\neq~0$ and $\psi_\mathrm{p}~\neq~0$ the equilibrium rotation period can yet be equal to the orbital period (see Fig.~\ref{fig:Gl581d_P_equ} for Gl581\,d, where $\omega_\mathrm{p}~=~n$ if $\psi_\mathrm{p}~\approx~74^\circ~\vee~106^\circ$). And in general, the rotation period of the planet will not be synchronized with respect to its orbit, if `synchronous' here means `equal' or `identical'. In general, we call $\mathrm{d}\omega_\mathrm{p}/\mathrm{d}t~=~0$ simply the state of `rotational equilibrium'. When rotational equilibrium meets $\omega_\mathrm{p}~\neq~n$, we call this constellation `pseudo-synchronous'. `Tidal locking' is used when there is no more transfer of angular momentum over the course of one orbit, i.e. $e~=~0$, $\psi_\mathrm{p}~=~0$, and $\omega_\mathrm{p}~=~n$. Strictly speaking, a body with $e~\neq~0$ or $\psi_\mathrm{p}~\neq~0$ cannot be tidally locked because $e$ and $\psi_\mathrm{p}$ evolve with time.

Uncertainties in our calculations arise from the assumption of a constant tidal dissipation in the planets, parametrized by the tidal dissipation factor $Q$ or $\tau$. For one thing, merging all the geophysical effects such as composition, viscosity, temperature distribution, and pressure in one parameter unquestionably is an oversimplification. In addition, it is unknown how $Q$ depends on the respective tidal frequency. Although on Earth $Q$ is constant over a a wide range of frequencies, the behavior will be different for different objects. The $Q$ value of terrestrial exoplanets remains to be estimated, either by observational constraints or theory. For another thing, whatever $Q$ value will turn out to describe terrestrial planets most reasonably, it will be a function of time due to the structural and rheological evolution of the planet.

Our results need to be considered against the background of further gravitational processes, such as the entourage of massive moons, the Kozai mechanism, Cassini states, planet-planet scattering, and inhomogeneities in the structure of the invoked bodies. These states and processes will add to -- and eventually counteract -- the tidal evolution. In those cases, significant obliquities could be maintained on time scales much larger or shorter than we present here.

\section{Conclusions}
\label{sec:conclusions}

Tidal processes raise severe constraints on the habitability of terrestrial planets in the IHZ of LMSs. First, tidal erosion of the obliquities, i.e. `tilt erosion', of such planets occurs much faster than life emerged on Earth. Planets with masses $\lesssim~10\,M_\mathrm{Earth}$ orbiting $\lesssim~0.25\,M_\odot$ stars in the IHZ lose initial Earth-like spins of $23.5^\circ$ within less than 0.1\,Gyr, typically much faster. In order to cover the complete IHZ with $t_\mathrm{ero.}~>~1$\,Gyr at high eccentricities, stellar masses $\gtrsim~0.9\,M_\odot$ are required. For terrestrial planets formed in highly eccentric orbits inside the IHZ, tilt erosion can occur within 10\,Gyr up to stellar masses of $1\,M_\odot$. 

Second, tidal heating of Earth-like planets in the IHZ of stars with $\lesssim~0.25\,M_\odot$ is significant. Depending on the body's obliquity, orbital semi-major axis, eccentricity, and its structural composition, a terrestrial planet in the IHZ of a $\lesssim~0.25\,M_\odot$ host star can undergo intense tidal heating with tidal surface heating rates $\gg~1$\,W/m$^2$. We find tidal surface heating rates of $2\,\mathrm{mW/m}^2$ to $15\,\mathrm{mW/m}^2$, depending on $\psi_\mathrm{p}$, for the Super-Earth candidate Gl581\,d and $<~150\,\mathrm{mW/m}^2$ for Gl581\,g. Simulations of coupled orbital-structural evolution of Earth-like planets are necessary to further explore the effect of tidal heating on habitability.

Third, the tidal equilibrium rotation period provided by Lec10 is a function of the planet's eccentricity and its obliquity. In general, as long as the planet maintains a significant obliquity or eccentricity, its rotational equilibrium period will not match the orbital period. As an example, we deduce the equilibrium rotation period of the extrasolar Super-Earth Gl581\,d to be around 35\,d. With an orbital period of roughly $68$\,d, this means $P_\mathrm{Gl581d}^\mathrm{equ.}~\approx~P_\mathrm{orb.}/2$. The model of FM08 neglects the impact of obliquity on the equilibrium rotation state.

For low eccentricities ($e~\lesssim~0.3$) and low obliquities ($\psi_\mathrm{p}~\lesssim~$ a few degrees), the CPL model of FM08 and the CTL model of Lec10 mathematically converge. In our calculations of $t_\mathrm{ero.}$ (Sect. \ref{sec:results}) they provide similar results in the low-eccentricity regime. Since the model of Lec10 considers higher terms in $e$, tidal processes in the moderate and high eccentricity regime occur on significantly shorter time scales than predicted by FM08. Hence, tidal heating computed with Lec10 is also more intense.

The possible detection of exomoons via transit photometry \citep{1999A&AS..134..553S}, the Rossiter-McLaughlin effect \citep{2010MNRAS.406.2038S}, planet-moon eclipses \citep{2007ASPC..366..242C}, and transit timing and duration variations \citep{2006A&A...450..395S,2009MNRAS.396.1797K} as well as the empirical constraint of the oblateness of a transiting planet \citep{2010ApJ...709.1219C,2010ApJ...716..850C}, are first steps towards the measurement of the obliquity of exoplanets. All these observational effects are only accessible for transiting planets. Since obliquity determines atmospheric conditions as well as the amount of tidal heating and the equilibrium rotation period, it will be indispensable to verify $\psi_\mathrm{p}$ for planets in the IHZ around their host stars to assess their habitability. Thus, transiting planets are the most promising targets for a comprehensive appraisal of an extrasolar habitat.

\begin{acknowledgements}
Our sincere thanks go to Brian Jackson for lively conversations about tidal processes. Virtual discussions with Sylvio Ferraz-Mello have been a valuable stimulation to this study. We appreciate discussions with D.~Breuer, H.~Rauer, F.~Sohl, H.~Hussmann, and T.~Spohn on a preliminary version of the manuscript. Thanks to the comments of the referee and the editor the understandability of this paper has been significantly improved. This study has benefited from inspirations during the AbGradCon2010 in T\"allberg, Sweden. For the most time of the paper processing, R.~Heller was supported by a PhD scholarship of the DFG Graduiertenkolleg 1351 ``Extrasolar Planets and their Host Stars'' at the Hamburger Sternwarte. The research leading to these results has received funding from the European Research Council under the European Community's Seventh Framework Programme (FP7/2007-2013 Grant Agreement no. 247060). R. Barnes acknowledges funding from NASA Astrobiology Institute's Virtual Planetary Laboratory lead team, supported by NASA under Cooperative Agreement No. NNH05ZDA001C. R. This research has made use of NASA's Astrophysics Data System Bibliographic Services.
\end{acknowledgements}

\bibliographystyle{aa} 
\bibliography{2010-2_Heller_et_al_Obliquity_Tides}

\clearpage

\clearpage

\clearpage

\clearpage

\clearpage

\clearpage

\end{document}